\documentclass[a4paper,11pt]{article}
\usepackage[a4paper,left=2.73cm,right=2.7cm,top=3cm,bottom=3.5cm]{geometry}

\usepackage{amsfonts,amsmath,amssymb,tabstackengine}
\usepackage{tikz}
\usepackage{array}

\definecolor{Gray}{gray}{0.95}

\usepackage[colorlinks=true,linktocpage=true,linkcolor=blue,citecolor=blue]{hyperref}
\usepackage{graphicx}

\addtocontents{toc}{\protect\setcounter{tocdepth}{2}}
\numberwithin{equation}{section}

 \usepackage [latin1]{inputenc}

\begin{document}

\begin{titlepage}

\thispagestyle{empty}

\begin{center}

{\LARGE \textbf{$\,\mathcal{N}=2\,$ supersymmetric S-folds}}

\vspace{40pt}
		
{\large \bf Adolfo Guarino}$\,^{a, b}$ \,\,\,  ,  \,\,\, {\large \bf Colin Sterckx}$\,^c$  \,\,\,  \large{and} \,\,\, {\large \bf Mario Trigiante}$\,^{d}$
		
\vspace{25pt}
		
$^a$\,{\normalsize
Departamento de F\'isica, Universidad de Oviedo,\\
Avda. Federico Garc\'ia Lorca 18, 33007 Oviedo, Spain.}
\\[7mm]

$^b$\,{\normalsize
Instituto Universitario de Ciencias y Tecnolog\'ias Espaciales de Asturias (ICTEA) \\
Calle de la Independencia 13, 33004 Oviedo, Spain.}
\\[7mm]

$^c$\,{\normalsize
Universit\'e Libre de Bruxelles (ULB) and International Solvay Institutes,\\
Service  de Physique Th\'eorique et Math\'ematique, \\
Campus de la Plaine, CP 231, B-1050, Brussels, Belgium.}
\\[7mm]

$^d$\,{\normalsize
Dipartimento di Fisica, Politecnico di Torino, Corso Duca degli Abruzzi 24,\\
I-10129 Turin, Italy and INFN, Sezione di Torino, Italy.}
\\[10mm]

\texttt{adolfo.guarino@uniovi.es} \,\, , \,\, \texttt{colin.sterckx@ulb.ac.be}  \,\, , \,\, \texttt{mario.trigiante@polito.it}

\vspace{20pt}

\vspace{20pt}
				
\abstract{
\noindent Multi-parametric families of AdS$_{4}$ vacua with various amounts of supersymmetry and residual gauge symmetry are found in the $\,[\textrm{SO}(1,1) \times \textrm{SO}(6)] \ltimes \mathbb{R}^{12}\,$ maximal supergravity that arises from the reduction of type IIB supergravity on $\,\mathbb{R}   \times  \textrm{S}^5\,$. These provide natural candidates to holographically describe new strongly coupled three-dimensional CFT's which are localised on interfaces of $\,\mathcal{N}=4\,$ super-Yang--Mills theory. One such AdS$_{4}$ vacua features a symmetry enhancement to $\,\textrm{SU}(2) \times \textrm{U}(1)\,$ while preserving $\,\mathcal{N}=2\,$ supersymmetry. Fetching techniques from the $\,\textrm{E}_{7(7)}\,$ exceptional field theory, its uplift to a class of $\,\mathcal{N}=2\,$ S-folds of type IIB supergravity of the form $\,\textrm{AdS}_{4} \times \textrm{S}^{1}   \times  \textrm{S}^5\,$ involving S-duality twists of hyperbolic type along $\,\textrm{S}^{1}\,$ is presented.}

\end{center}

\end{titlepage}

\tableofcontents

\hrulefill
\vspace{10pt}

\section{Introduction}

Electromagnetic duality in four-dimensional maximal supergravity has {provided} a very rich phenomenology as far as the existence of new gaugings and vacuum solutions are concerned. The prototypical example is the dyonically-gauged $\,\textrm{SO}(8)\,$ supergravity where the action of electromagnetic duality on the gauging generates a one-parameter family of inequivalent theories parameterised by a continuous parameter $\,c \in \left[  0 \, , \, \sqrt{2} - 1 \right]  \,$ \cite{Dall'Agata:2012bb}. Setting the parameter to $\,c=0\,$ then the standard (electric) $\,\textrm{SO}(8)\,$ supergravity of de Wit and Nicolai \cite{deWit:1982ig} is recovered which is known to arise upon dimensional reduction of eleven-dimensional supergravity on a seven-sphere $\,\textrm{S}^7\,$. The various AdS$_{4}$ vacua of the $\,c=0\,$ theory \cite{Warner:1983vz} (see also \cite{Comsa:2019rcz} for an updated encyclopedic 	reference) get generalised to one-parameter families of vacua when turning on $\,c\,$ and, more importantly, new and genuinely dyonic AdS$_{4}$ vacua also appear which do not have a well defined (electric) $\,c \rightarrow 0\,$ limit \cite{Dall'Agata:2012bb,Borghese:2012qm,Borghese:2012zs,Borghese:2013dja}. Other types of four-dimensional solutions, like domain-walls \cite{Guarino:2013gsa,Tarrio:2013qga} or black holes \cite{Anabalon:2013eaa,Lu:2014fpa,Wu:2015ska}, have also been investigated using instead a phase-like parameterisation $\,\omega = \arg(1+i \, c) \in \left[  0 \, , \, {\pi/8 }\right]\,$ of the electromagnetic deformation parameter. However, and despite the rich structure of new solutions at $\,c \neq 0\,$, the question about the eleven-dimensional interpretation of the electromagnetic parameter $\,c\,$ remains elusive and various no-go theorems have been stated against the existence of such a higher dimensional origin \cite{deWit:2013ija,Lee:2015xga}.  Also, for the new supersymmetric AdS$_4$ vacua at $\,c \neq 0\,$, the holographic interpretation of the deformation parameter remains obscure from the perspective of the AdS$_{4}$/CFT$_{3}$ correspondence.

Unlike for the $\,\textrm{SO}(8)\,$ theory, much more is by now known about the dyonically-gauged $\,\textrm{ISO}(7)\,$ supergravity that arises from the reduction of massive IIA supergravity on a six-sphere $\,\textrm{S}^{6}\,$ \cite{Guarino:2015vca}. In this case the electromagnetic deformation parameter is a discrete (on/off) deformation, namely, it can be set to $\,c=0 \textrm{ or } 1\,$ without loss of generality \cite{Dall'Agata:2014ita}. Various AdS$_{4}$ \cite{Guarino:2015qaa,Guarino:2019jef,Guarino:2019snw}, domain-wall \cite{Guarino:2016ynd,Guarino:2019snw}, and black hole \cite{Guarino:2017eag,Guarino:2017pkw,Hosseini:2017fjo,Benini:2017oxt} solutions have been constructed which necessarily require a non-zero electromagnetic deformation parameter $\,c\,$. Within this massive IIA context, the electromagnetic parameter is identified with the Romans mass parameter $\,\hat{F}_{0}\,$ of the ten-dimensional theory \cite{Romans:1985tz}, and has a holographic interpretation as the Chern--Simons level $\,k\,$ of a three-dimensional super-Chern--Simons dual theory \cite{Guarino:2015jca}.

\begin{figure}[h]
\label{Figure_diagram}
\vspace{2.5mm}
\centering
\begin{tikzpicture}
	\node at(-3,4){$D=10$};
	\node at(-3,1.5){$D=4$};
	\node at(-3,-0.2){$D=3$};
	\node[draw,text width=9cm,text centered] at(4.8,4){\mbox{Type IIB \, \& \, S-fold with $\,\textrm{AdS}_{4} \times \textrm{S}^1 \times \textrm{S}^5 $ geometry}};
		\node[draw,text width=5.4cm,text centered] at(1.6,1.5){$\,[\textrm{SO}(1,1) \times \textrm{SO}(6)] \ltimes \mathbb{R}^{12}\,$\\ \mbox{gauging with an AdS$_{4}$ vacuum}};
\node[draw,text width=4.5cm,text centered] at(7.5,1.5){\mbox{$\mathcal{N}=4$ SYM} \mbox{with a localised interface}};
	\node[draw,text width=2.3cm,text centered] at(6.6,-0.5){\mbox{$J$-fold CFT$_{3}$}};
	\draw[<-](1.2,2.2)--(1.2,3.6);
	\node[text width=1.8cm,text justified] at(0.1,2.8){Reduction on $\,\mathbb{R} \times \textrm{S}^5$};
  \draw[->](3.3,2.2)--(3.3,3.6);
	\node[text width=1.8cm,text justified] at(4.6,2.8){\mbox{Uplift method : $\textrm{E}_{7(7)}$-EFT involving} \mbox{hyperbolic twists $\,A_{(k)}\,$ along $\,\textrm{S}^1\,$}};
	\draw[<->](5.2,-0.2)--(2,0.8);
	\node[text width=3.2cm,text justified,rotate=-18] at(4,-0.2){AdS$_{4}$/CFT$_{3}$};
	\draw[->](5.8,0.8)--(5.8,-0.1);
	\node[text width=2.5cm, text justified] at(7.5,0.4){\mbox{$J \in \textrm{SL}(2,\mathbb{Z})_{\textrm{IIB}}$ action}};
\end{tikzpicture}
\caption{Type IIB  S-folds with hyperbolic monodromies $\,\mathfrak{M}(k) = - \mathcal{S} \, \mathcal{T}^{k}\,$ along $\,\textrm{S}^1\,$ and connection with three-dimensional $J$-fold CFT's.}
\vspace{2.5mm}
\end{figure}
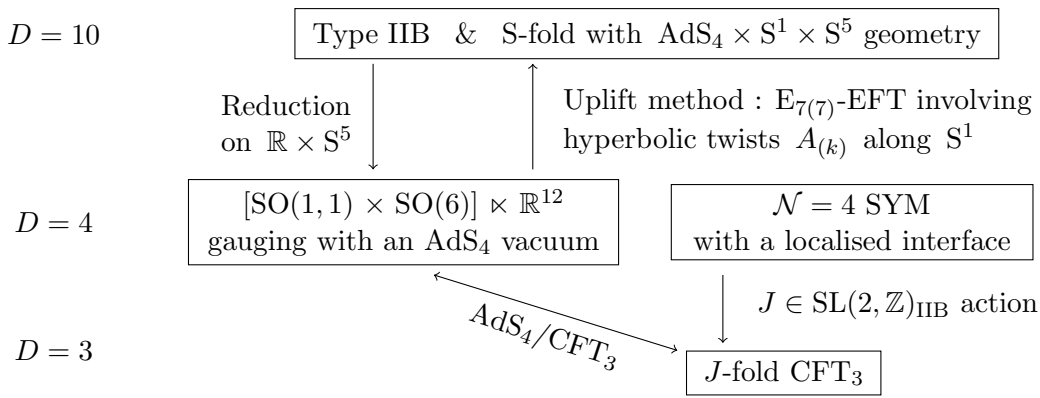

The role of the electromagnetic deformation $\,c\,$ has been much less investigated  in the context of type IIB supergravity. The relevant dyonically-gauged supergravity in this case is the $\,{[\textrm{SO}(1,1) \times \textrm{SO}(6)] \ltimes \mathbb{R}^{12}}\,$ theory which arises from the reduction of type IIB supergravity on the product $\,\mathbb{R} \times \textrm{S}^{5}\,$ \cite{Inverso:2016eet}.  As for the ISO(7) theory, the electromagnetic deformation is again a discrete (on/off) deformation, namely, $\,c=0 \textrm{ or } 1\,$ \cite{Dall'Agata:2014ita}. This four-dimensional supergravity has been shown to contain various types of AdS$_{4}$ vacua preserving different amounts of supersymmetry as well as of residual gauge symmetry. In particular, an $\,{\mathcal{N}=4}\,$ and $\,\textrm{SO}(4)\,$ symmetric solution was reported in \cite{Gallerati:2014xra} and subsequently, in \cite{Inverso:2016eet}, uplifted to a class of $\,\textrm{AdS}_{4} \times \textrm{S}^{1} \times \textrm{S}^{5}\,$ S-fold backgrounds of type IIB supergravity using the $\,\textrm{E}_{7(7)}\,$ exceptional field theory ($\textrm{E}_{7(7)}$-EFT). These S-folds involve S-duality twists $\,A_{(k)}\,$ ($k \ge 3$) that induce $\,\textrm{SL}(2,\mathbb{Z})_{\textrm{IIB}}\,$ monodromies $\,\mathfrak{M}(k) = - \mathcal{S} \, \mathcal{T}^{k}\,$  of hyperbolic type along $\,\textrm{S}^1\,$, and can be systematically constructed as quotients of degenerate Janus-like solutions of the type IIB theory \cite{Bak:2003jk,DHoker:2006vfr} where the string coupling $\,g_{s}\,$ diverges at infinity. Together with the $\,\mathcal{N}=4 \,\, \& \,\,\textrm{SO}(4)\,$ solution, additional $\,{\mathcal{N}=0 \,\, \& \,\,\textrm{SO}(6)}\,$ \cite{DallAgata:2011aa} and $\,\mathcal{N}=1 \,\, \& \,\,\textrm{SU}(3)\,$ \cite{Guarino:2019oct} solutions have been found and uplifted to similar S-fold backgrounds of type IIB supergravity with hyperbolic monodromies in \cite{Guarino:2019oct}. From a holographic perspective, these AdS$_{4}$ vacua describe new strongly coupled three-dimensional CFT's, referred to as $J$-fold CFT's in \cite{Assel:2018vtq} (see also \cite{Garozzo:2018kra,Garozzo:2019ejm} and \cite{Bobev:2019jbi}), which are localised on interfaces of $\,{\mathcal{N}=4}\,$ super-Yang--Mills theory (SYM) \cite{Clark:2004sb}. In the $\,\mathcal{N}=4\,$ case \cite{Assel:2018vtq}, a hyperbolic monodromy $\, J = - \mathcal{S} \, \mathcal{T}^{k} \in \textrm{SL}(2,\mathbb{Z})_{\textrm{IIB}}\,$ was shown to introduce a Chern--Simons level $\,k\,$ in the dual $J$-fold CFT which, in turn, is constructed from the $\,T(U(N))\,$ theory \cite{Gaiotto:2008ak} upon suitable gauging of flavour symmetries. A diagram illustrating this type IIB construction is depicted in Figure~\ref{Figure_diagram}.

On the other hand, a classification of interface SYM theories was performed in \cite{DHoker:2006qeo} (see also \cite{Gaiotto:2008sd}) in correspondence to the various amounts of supersymmetry, as well as the largest possible global symmetry, preserved by the interface operators. Three supersymmetric cases were identified: interfaces with $\,\mathcal{N}=4 \,\, \& \,\,\textrm{SO}(4)\,$ symmetry, $\,{\mathcal{N}=2 \,\, \& \,\,\textrm{SU}(2) \times \textrm{U}(1)}\,$ symmetry and $\,\mathcal{N}=1 \,\, \& \,\,\textrm{SU}(3)\,$ symmetry. While the S-folds in \cite{Inverso:2016eet} and \cite{Guarino:2019oct} respectively match the symmetries of the $\,\mathcal{N}=4\,$ and $\,\mathcal{N}=1\,$ cases, the gravity duals of the would be $\,\mathcal{N}=2\,$ $J$-fold CFT's localised on the interface with $\,\textrm{SU}(2) \times \textrm{U}(1)\,$ symmetry remain missing. In this work we fill this gap and present a new family of $\,\textrm{AdS}_{4} \times \textrm{S}^{1} \times \textrm{S}^{5}\,$ S-folds with $\,\mathcal{N}=2\,$ supersymmetry, $\,\textrm{SU}(2) \times \textrm{U}(1)\,$ symmetry and, as in the previous cases, involving S-duality twists that induce monodromies of hyperbolic type along $\,\textrm{S}^1\,$.

The paper is organised as follows. In Section~\ref{Sec:4D} we perform a study of multi-parametric families of AdS$_{4}$ vacua in the $\,{[\textrm{SO}(1,1) \times \textrm{SO}(6)] \ltimes \mathbb{R}^{12}}\,$ maximal supergravity. We find four families of vacua, one of them being $\,\mathcal{N}=2\,$ supersymmetric and containing a vacuum with a residual symmetry enhancement to $\,\textrm{SU}(2) \times \textrm{U}(1)\,$. In Section~\ref{sec:10D}, by implementing a generalised Scherk--Schwarz (S--S) ansatz in $\textrm{E}_{7(7)}$-EFT, we uplift such an AdS$_{4}$ vacuum to a class of $\,\textrm{AdS}_{4} \times \textrm{S}^{1} \times \textrm{S}^{5}\,$ $\,\mathcal{N}=2 \,$ S-folds of type IIB supergravity with $\,\textrm{SU}(2) \times \textrm{U}(1)\,$ symmetry and a non-trivial hyperbolic monodromy along $\,\textrm{S}^1\,$. In Section~\ref{Sec:conclusions} we present our conclusions and discuss future directions.

\section{AdS$_{4}$ vacua of $\,{[\textrm{SO}(1,1) \times \textrm{SO}(6)] \ltimes \mathbb{R}^{12}}\,$ maximal supergravity}
\label{Sec:4D}

We continue the study of AdS$_{4}$ vacua initiated in \cite{DallAgata:2011aa}, and further investigated in \cite{Gallerati:2014xra} and \cite{Guarino:2019oct}, for the dyonically-gauged maximal supergravity with non-abelian gauge group
\begin{equation}
\label{G-group}
\textrm{G}={[\textrm{SO}(1,1) \times \textrm{SO}(6)] \ltimes \mathbb{R}^{12}}  \ .
\end{equation}
We will show how the AdS$_{4}$ vacua of \cite{DallAgata:2011aa,Gallerati:2014xra,Guarino:2019oct} actually correspond to very special points (featuring residual symmetry enhancements) within multi-parametric families of solutions. Each of these families preserves a given amount supersymmetry, namely, $\,\mathcal{N}=0,\, 1,\, 2 \textrm{ or } 4\,$. More specifically we find:
\begin{itemize}

\item  A three-parameter family of $\,\mathcal{N}=0  \,\, \& \,\, \textrm{U}(1)^3\,$ symmetric AdS$_{4}$ vacua with symmetry enhancements to $\,\textrm{SU}(2) \times \textrm{U}(1)^2\,$,  $\,\textrm{SU}(3) \times \textrm{U}(1)\,$ and  $\,\textrm{SO}(6) \sim \textrm{SU}(4)\,$ at specific values of the three arbitrary parameters.

\item  A two-parameter family of $\,\mathcal{N}=1 \,\, \& \,\, \textrm{U}(1)^2\,$ symmetric AdS$_{4}$ vacua with symmetry enhancements to $\,\textrm{SU}(2) \times \textrm{U}(1)\,$ and  $\,\textrm{SU}(3)\,$ at specific values of the two arbitrary parameters.

\item  A one-parameter family of $\,\mathcal{N}=2 \,\, \& \,\, \textrm{U}(1)^2\,$ symmetric AdS$_{4}$ vacua with a symmetry enhancement to $\,\textrm{SU}(2) \times \textrm{U}(1)\,$ at a special value of the arbitrary parameter.

\item  A single $\,\mathcal{N}=4 \,\, \& \,\, \textrm{SO}(4)\,$ symmetric AdS$_{4}$ vacuum.

\end{itemize}

The $\,\mathcal{N}=2\,$ family of AdS$_{4}$ vacua is new and we will uplift the solution with $\,\textrm{SU}(2) \times \textrm{U}(1)\,$ enhanced residual symmetry to a new and analytic family of S-fold backgrounds of type IIB supergravity in Section~\ref{sec:10D}.

\subsection{The $\,\mathcal{N}=8\,$ theory: gauging and scalar potential}

We follow the conventions and notation of \cite{Guarino:2019oct}, which slightly differ from those of \cite{Inverso:2016eet}, to describe the dyonically-gauged maximal supergravity with gauge group $\,\textrm{G}\,$ in (\ref{G-group}). For the purposes of this work, \textit{i.e.} the study of $\,\textrm{AdS}_4\,$ vacua, we set to zero all the vector and (auxiliary \cite{deWit:2005ub}) tensor fields of the theory, so that the bosonic Lagrangian reduces to the following one 
\begin{equation}
\label{Lagrangian}
\mathcal{L}_{\mathcal{N}=8} = \left(  \frac{R}{2} - V_{\mathcal{N}=8} \right)  * 1 + \frac{1}{96} \, \textrm{Tr}\left(  dM \wedge * dM^{-1}\right) \,,
\end{equation}
which describes the scalar fields $\,M_{MN}\,$ coupled to Einstein gravity in the presence of a scalar potential. The scalar fields serve as coordinates on the coset space of maximal supergravity
\begin{equation}
\label{M_scalar}
M_{MN} = \mathcal{V} \, \mathcal{V}^{t}  \in \frac{\textrm{E}_{7(7)}}{\textrm{SU}(8)} \ ,
\end{equation}
with $\,M=1,\ldots,56\,$ being a fundamental index of $\,\textrm{E}_{7(7)}\,$. The coset representative $\,\mathcal{V}\,$ is constructed by direct exponentiation of the $\,70\,$ non-compact generators $\,t_{A}{}^{B}\,$ (with $\,{t_{A}{}^{A}=0}\,$) and $\,t_{ABCD}=t_{[ABCD]}\,$ generators of $\,\textrm{E}_{7(7)}\,$ in the SL(8) basis\footnote{We adopt the conventions in the appendix of \cite{Guarino:2015tja} for the explicit form of the $\,t_{A}{}^{B}\,$ and $\,t_{ABCD}\,$ matrices.}. The scalar potential in (\ref{Lagrangian}), which survives our truncation to the Einstein-scalar sector, is induced by the gauging of the group $\,\textrm{G}\,$ in (\ref{G-group}) within the maximal theory and has the following general form:
\begin{equation}
\label{V_N8}
V_{\mathcal{N}=8} =  \frac{g^{2}}{672} \,  {X_{MN}}^{R} \,  {X_{PQ}}^{S} \, M^{MP} \left( M^{NQ}  \, M_{RS} +   7 \,  \delta^{Q}_{R} \, \delta^{N}_{S}  \right) \ ,
\end{equation}
which depends on the gauge coupling $\,g\,$, the scalar matrix $\,M_{MN}\,$ (and its inverse $\,M^{MN}\,$) and on a constant \textit{embedding tensor} $\,X_{MN}{}^{P}\,$ living in the $\,\textbf{912}\,$ of $\,\textrm{E}_{7(7)}\,$ \cite{deWit:2007mt}. This tensor codifies how the gauge group $\,\textrm{G}\,$ is embedded into the $\,\textrm{E}_{7(7)}\,$ duality group of maximal supergravity. Moreover,  it also specifies the gauge connection which involves both electric and magnetic vector fields transforming under the $\,\textrm{Sp}(56)\,$ group of electromagnetic transformations of the theory (for reviews see \cite{Samtleben:2008pe,Trigiante:2016mnt}).

Under $\,\textrm{SL}(8) \subset \textrm{E}_{7(7)}\,$ the index $\,M\,$ decomposes into antisymmetric pairs  $\,_M=(_{[AB]},^{[AB]})\,$ where $\,A=1,\ldots,8\,$ denotes a fundamental index of $\,\textrm{SL}(8)\,$. This implies that, for gaugings of subgroups of $\,\textrm{SL}(8)$, the non-vanishing electric and magnetic components of the embedding tensor are given by \cite{DallAgata:2011aa}
\begin{equation}
\label{X_tensor}
\begin{array}{llllllll}
\textrm{electric } & : & \hspace{3mm}  {X_{[AB] [CD]}}^{[EF]} &=& - X_{[AB] \phantom{[EF]} [CD]}^{\phantom{[AB]} [EF]} &=& -8 \, \delta_{[A}^{[E} \, \eta_{B][C} \, \delta_{D]}^{F]}  &  , \\[4mm]
\textrm{magnetic } & :  & \hspace{3mm}  X^{[AB] \phantom{[CD]}[EF]}_{\phantom{[AB]}[CD]} &=& - {X^{[AB] [EF]}}_{[CD]} &=& -8 \, \delta_{[C}^{[A} \, \tilde{\eta}^{B][E} \, \delta_{D]}^{F]} &  ,
\end{array}
\end{equation}
in terms of two symmetric matrices $\,\eta_{AB}\,$ and $\,\tilde{\eta}^{AB}\,$.  For the gauging of $\,\textrm{G} \subset \textrm{SL}(8)\,$ in (\ref{G-group}) these are
\begin{equation}
\eta_{AB} = \textrm{diag} (\, 0 \, , \, \mathbb{I}_{6}\, ,  \, 0 \,)
\hspace{10mm} \textrm{ and } \hspace{10mm}
\tilde{\eta}^{AB} = c \,\,  \textrm{diag} (\, -1 \, , \, 0_{6}\, ,  \, 1 \,) \ .
\end{equation}
As stated in the introduction, the magnetic part of the embedding tensor in (\ref{X_tensor}) allows for an (on/off) electromagnetic parameter $\,c\,$ so that $\,\tilde{\eta}^{AB} \propto c\,$.


\subsection{$\,\mathbb{Z}_{2}^{3}\,$ invariant sector}
\label{sec:N=1_model}

In order to efficiently search for extrema of the scalar potential (\ref{V_N8}), we will now construct a $\,\mathbb{Z}_{2}^{3}\,$ invariant sector of the $\,{[\textrm{SO}(1,1) \times \textrm{SO}(6)] \ltimes \mathbb{R}^{12}} \,$ maximal supergravity. This sector can be recast as a minimal $\,\mathcal{N}=1\,$ supergravity coupled to seven chiral multiplets $\,z_{i}\,$ with $\,{i=1,\ldots,7}\,$. The same invariant sector has recently been explored in the dyonically-gauged $\,\textrm{ISO}(7)\,$ theory \cite{Guarino:2019snw} and the purely electric $\,\textrm{SO}(8)\,$ theory \cite{Bobev:2019dik}, and it originally appeared in the context of type II orientifold compactifications with generalised fluxes \cite{Aldazabal:2006up,Aldazabal:2008zza}.

To describe this sector of the maximal theory, we first focus on a four-element Klein subgroup of $\,\textrm{SL}(8)\,$. Its action on the fundamental index $\,A\,$ is given by
\begin{equation}
\label{Z2xZ2xZ2_action}
\begin{array}{rl}
\mathbb{Z}^{(1)}_{2} : &  (x_{1}\,,\,x_{2}\,,\,x_{3}\,,\,x_{4}\,,\,x_{5}\,,\,x_{6}\,,\,x_{7}\,,\,x_{8}) \rightarrow  (x_{1}\,,\,x_{2}\,,\,x_{3}\,,\,-x_{4}\,,\,-x_{5}\,,\,-x_{6}\,,\,-x_{7}\,,\,x_{8}) \ ,\\[2mm]
\mathbb{Z}^{(2)}_{2}  : &  (x_{1}\,,\,x_{2}\,,\,x_{3}\,,\,x_{4}\,,\,x_{5}\,,\,x_{6}\,,\,x_{7}\,,\,x_{8}) \rightarrow  (x_{1}\,,\,-x_{2}\,,\,-x_{3}\,,\,x_{4}\,,\,x_{5}\,,\,-x_{6}\,,\,-x_{7}\,,\,x_{8}) \ ,
\end{array}
\end{equation}
together with the remaining generators $\,\mathbb{I}\,$ and $\,\mathbb{Z}^{(1)}_{2} \, \mathbb{Z}^{(2)}_{2}\,$. In addition, we will also require invariance under an extra $\,\mathbb{Z}^{*}_{2}\,$ generator acting as
\begin{equation}
\begin{array}{rl}
\mathbb{Z}^{*}_{2} : &  (x_{1}\,,\,x_{2}\,,\,x_{3}\,,\,x_{4}\,,\,x_{5}\,,\,x_{6}\,,\,x_{7}\,,\,x_{8}) \rightarrow  (x_{1}\,,\,-x_{2}\,,\,x_{3}\,,\,-x_{4}\,,\,x_{5}\,,\,-x_{6}\,,\,x_{7}\,,\,-x_{8}) \ .
\end{array}
\end{equation}
The resulting $\,\mathbb{Z}_{2}^{3}\,$ invariant sector describes $\,\mathcal{N}=1\,$ supergravity coupled to seven chiral multiplets (and no vector multiplets)
\begin{equation}
\label{Phi_def_7}
z_{i} = - \chi_{i} + i \, e^{-\varphi_{i}}
\hspace{10mm} \textrm{ with } \hspace{10mm} i=1, \ldots, 7 \ .
\end{equation}
The fourteen real spinless fields are associated with generators $\,t_{A}{}^{B}\,$ (scalars) and $\,t_{[ABCD]}\,$ (pseudo-scalars) of $\,\textrm{E}_{7(7)}\,$ in the SL(8) basis. The former have associated generators of the form
\begin{equation}
\label{generators_scalars}
\begin{array}{llll}
g_{\varphi_{1}}  & = & -t_{1}{}^{1} -  t_{2}{}^{2} -  t_{3}{}^{3} +  t_{4}{}^{4} +  t_{5}{}^{5} +  t_{6}{}^{6} +  t_{7}{}^{7} -  t_{8}{}^{8} & , \\[2mm]
g_{\varphi_{2}}  & = & -t_{1}{}^{1} +  t_{2}{}^{2} +  t_{3}{}^{3} -  t_{4}{}^{4} -  t_{5}{}^{5} +  t_{6}{}^{6} +  t_{7}{}^{7} -  t_{8}{}^{8} & , \\[2mm]
g_{\varphi_{3}}  & = & -t_{1}{}^{1} +  t_{2}{}^{2} +  t_{3}{}^{3} +  t_{4}{}^{4} +  t_{5}{}^{5} -  t_{6}{}^{6} -  t_{7}{}^{7} -  t_{8}{}^{8} & , \\[4mm]
g_{\varphi_{4}}  & = & t_{1}{}^{1} -  t_{2}{}^{2} +  t_{3}{}^{3} +  t_{4}{}^{4} -  t_{5}{}^{5} +  t_{6}{}^{6} -  t_{7}{}^{7} -  t_{8}{}^{8} & , \\[2mm]
g_{\varphi_{5}}  & = & t_{1}{}^{1} +  t_{2}{}^{2} -  t_{3}{}^{3} -  t_{4}{}^{4} +  t_{5}{}^{5} +  t_{6}{}^{6} -  t_{7}{}^{7} -  t_{8}{}^{8} & , \\[2mm]
g_{\varphi_{6}}  & = & t_{1}{}^{1} +  t_{2}{}^{2} -  t_{3}{}^{3} +  t_{4}{}^{4} -  t_{5}{}^{5} -  t_{6}{}^{6} +  t_{7}{}^{7} -  t_{8}{}^{8} & , \\[4mm]
g_{\varphi_{7}}  & = & t_{1}{}^{1} -  t_{2}{}^{2} +  t_{3}{}^{3} -  t_{4}{}^{4} +  t_{5}{}^{5} -  t_{6}{}^{6} +  t_{7}{}^{7} -  t_{8}{}^{8} & ,
\end{array}
\end{equation}
whereas the latter correspond with generators given by
\begin{equation}
\label{generators_pseudo-scalars}
\begin{array}{lllllllllll}
g_{\chi_{1}}  & = &   t_{1238}   & \hspace{5mm},\hspace{5mm}  & \,\,\,  g_{\chi_{4}}  & = &    t_{2578}  & \hspace{5mm},\hspace{5mm} \\[2mm]
g_{\chi_{2}}  & = &    t_{1458}  & \hspace{5mm},\hspace{5mm} & \,\,\, g_{\chi_{5}}  & = &    t_{4738}  & \hspace{5mm},\hspace{5mm} & \,\,\,\, g_{\chi_{7}}  & = &    t_{8246}  \  .\\[2mm]
g_{\chi_{3}}  & = &    t_{1678}  & \hspace{5mm},\hspace{5mm} & \,\,\, g_{\chi_{6}}  & = &    t_{6358}  & \hspace{5mm},\hspace{5mm}
\end{array}
\end{equation}
Exponentiating (\ref{generators_scalars}) and (\ref{generators_pseudo-scalars}) with coefficients $\,\varphi_{i}\,$ and $\,\chi_{i}\,$ as
\begin{equation}
\label{M_scalar_N=1}
\mathcal{V} = \textrm{Exp} \left[ {-12 \, \displaystyle\sum_{i=1}^7 \chi_{i}\, g_{\chi_{i}}} \right]
\,
\textrm{Exp} \left[ {\frac{1}{4} \, \displaystyle\sum_{i=1}^7 \varphi_{i}\, g_{\varphi_{i}}} \right] \ ,
\end{equation}
yields a parameterisation of an $\,M_{MN}= \mathcal{V} \, \mathcal{V}^{t} \in [\textrm{SL}(2)/\textrm{SO}(2)]^7\,$ subspace of the coset space in (\ref{M_scalar}). The kinetic terms in the resulting $\,\mathcal{N}=1\,$ sector follow from (\ref{Lagrangian}) and (\ref{M_scalar_N=1}), and are given by
\begin{equation}
\mathcal{L}_{kin}=-\frac{1}{4} \displaystyle\sum_{i=1}^7  \left[  (\partial \varphi_{i})^2 + e^{2 \varphi_{i}} \,  (\partial \chi_{i})^2 \right] \ .
\end{equation}
These match the standard kinetic terms $\,\mathcal{L}_{kin}=- (\partial^{2}_{z_{i} , \bar{z}_{j}} K )\, dz_{i} \wedge * d\bar{z}_{j}\,$ for a set of seven chiral fields $\,z_{i}\,$ with K\"ahler potential
\begin{equation}
\label{K_7_chirals}
K = - \displaystyle\sum_{i=1}^7   \log[-i(z_{i}-\bar{z}_{i})] \ .
\end{equation}
Lastly, when restricted to the $\,\mathbb{Z}_{2}^{3}\,$ invariant sector entering (\ref{M_scalar_N=1}), the scalar potential, as computed from (\ref{V_N8}), can be recovered from a holomorphic superpotential
\begin{equation}
\label{W_Z2xZ2}
W = 2 \, g \, \big[  \, z_{1} z_{5} z_{6} +  z_{2} z_{4} z_{6} + z_{3} z_{4} z_{5} + (  z_{1} z_{4} +  z_{2} z_{5} + z_{3} z_{6} ) \,  z_{7} \, \big] + 2 \,  g \, c  \, ( 1 - z_{4}  z_{5} z_{6}  z_{7} )  \ ,
\end{equation}
using the standard $\,\mathcal{N}=1\,$  formula
\begin{equation}
V_{\mathcal{N}=1} = e^{K} \Big[ K^{z_{i} \bar{z}_{j}}  \, D_{z_{i}} W  \, D_{\bar{z}_{j}} \bar{W}   - 3 \, W \, \bar{W}  \Big] \ ,
\end{equation}
where $\,D_{z_{i}} W \equiv \partial_{z_{i}} W + (\partial_{z_{i}} K) W\,$ is the K\"ahler derivative and $\,K^{z_{i} \bar{z}_{j}} \,$ is the inverse of the K\"ahler metric $\,K_{z_{i} \bar{z}_{j}} \equiv \partial^{2}_{z_{i} , \bar{z}_{j}} K \,$. Note that only the last term in the superpotential (\ref{W_Z2xZ2}) turns out to be sensitive to the electromagnetic parameter $\,c\,$.

\subsection{New families of AdS$_4$ vacua}
\label{sec:AdS4_vacua}

A thorough study of the structure of extrema of the scalar potential (\ref{V_N8}), restricted to the  $\,\mathbb{Z}_{2}^{3}\,$ invariant sector, reveals a rich structure of (fairly) symmetric AdS$_4$ vacua. We find four families of vacua preserving $\,{\mathcal{N}=0,\, 1,\, 2 \textrm{ or } 4}\,$ supersymmetry as well as various residual gauge symmetries ranging from $\,\textrm{U}(1)^2\,$ to $\,\textrm{SO}(6) \sim \textrm{SU}(4)\,$. The three supersymmetric families are also supersymmetric within the $\,\mathcal{N}=1\,$ model with seven chirals presented in the previous section, and therefore satisfy the F-flatness conditions
\begin{equation}
D_{z_{i}} W = 0 \ ,
\end{equation}
that follow from the superpotential (\ref{W_Z2xZ2}) and K\"ahler potential (\ref{K_7_chirals}). Importantly, all the AdS$_4$ vacua we will present in this section are genuinely dyonic, namely, they disappear if taking the limit $\,c \rightarrow 0\,$ to a purely electric gauging of $\,\textrm{G}\,$ in (\ref{G-group}).


\subsubsection{$\mathcal{N}=0\,$ vacua with $\,\textrm{U}(1)^3 \rightarrow \textrm{SU}(2) \times \textrm{U}(1)^2  \rightarrow \textrm{SU}(3) \times \textrm{U}(1)  \rightarrow \textrm{SO}(6) \,$ symmetry}

There is a three-parameter family of $\,\mathcal{N}=0\,$ solutions that preserves $\,\textrm{U}(1)^3\,$ and is located at
\begin{equation}
\label{VEVs_z_N0}
z_{1,2,3} = c \left( - \chi_{1,2,3} +  i \, \frac{1}{\sqrt{2}} \right)
\hspace{10mm} \textrm{and} \hspace{10mm}
z_{4}=z_{5}=z_{6}=z_{7}= i \ ,
\end{equation}
with $\,\chi_{1,2,3}\,$ being arbitrary (real) parameters. This family of solutions has a vacuum energy given by
\begin{equation}
V_{0} = -2 \,  \sqrt{2}  \,\, g^2 \, c^{-1} \ ,
\end{equation}
and a spectrum of normalised scalar masses of the form
\begin{equation}
\label{spectrum_SO6_scalars}
\begin{array}{lll}
m^2 L^2 &=&
 6 \,\,\, ( \times 2 )
\hspace{3mm} , \hspace{3mm}
 -3 \,\,\, ( \times 2 )
 \hspace{3mm} , \hspace{3mm}
 0 \,\,\, ( \times 28  ) \ , \\[4mm]
 & &
  - \frac{3}{4}  + \frac{3}{2}  \, \chi^2    \,\,\, ( \times  2 )  \ ,  \\[4mm]
 & &
 - \frac{3}{4}  + \frac{3}{2}  \left(  \chi - 2\chi_ i \right){}^2   \,\,\, ( \times  2 )     \hspace{8mm} i=1,2,3  \ , \\[4mm]
 & &
 - \frac{3}{4}  + \frac{3}{2}  \chi_{i}^2   \,\,\, ( \times  4 )     \hspace{8mm} i=1,2,3  \ , \\[4mm]
 & &
 -3 + 6 \chi_i^2  \,\,\, ( \times 2 )  \hspace{8mm} i=1,2,3  \ , \\[4mm]
 & &

 -3  + \frac{3}{2}  \, ( \chi_i \pm \chi_j )^2    \,\,\, ( \times  2 ) \hspace{8mm} i <  j  \ ,
\end{array}
\end{equation}
where $\,\chi \equiv \chi_1 + \chi_2 + \chi_3 \,$ and $\,L^2=-3/V_{0}\,$ is the AdS$_{4}$ radius. This family of solutions is perturbatively unstable due to the mass eigenvalue $\,-3\,$ lying below the Breitenlohner-Freedman bound for stability in AdS$_{4}$ \cite{Breitenlohner:1982bm} . The computation of the vector masses yields
\begin{equation}
\label{spectrum_SO6_vectors}
\begin{array}{lll}
m^2 L^2 &=&
0  \,\,\, ( \times 3 )
\hspace{3mm} , \hspace{3mm}
6  \,\,\, ( \times 1 ) \ , \\[4mm]
& &
\frac{9}{4}   +  \frac{3}{2} \, \chi _i^2    \,\,\, ( \times 4 )
\hspace{8mm} i=1,2,3 \ , \\[4mm]
& &
\frac{3}{2} \left(\chi _i \pm  \chi _j \right){}^2  \,\,\, ( \times 2 )
\hspace{8mm} i <  j  \ .
\end{array}
\end{equation}
Note that a generic solution in this family preserves a $\,\textrm{U}(1)^3\,$ symmetry as three vectors are generically massless. Therefore, out of the $\,28\,$ massless scalars in \eqref{spectrum_SO6_scalars}, only $\,3\,$ of them correspond to physical directions in the scalar potential. Imposing a pairwise identification between the free axions $\,\chi_{1,2,3}\,$ results in a symmetry enhancement to $\,\textrm{SU}(2) \times \textrm{U}(1)^2\,$. A further identification $\,\chi_{1}=\chi_{2}=\chi_{3} \neq 0\,$ implies a symmetry enhancement to $\,\textrm{SU}(3) \times \textrm{U}(1)\,$. Lastly, setting $\,\chi_{1,2,3}=0\,$ enhances the symmetry to $\,\textrm{SU}(4) \sim \textrm{SO}(6)\,$. This $\textrm{SO}(6)$ symmetric solution was originally studied in \cite{Bak:2003jk} from a ten-dimensional perspective and, more recently, connected with a family of type IIB S-fold backgrounds in \cite{Guarino:2019oct}.

\subsubsection{$\mathcal{N}=1\,$ vacua with $\,\textrm{U}(1)^2 \rightarrow \textrm{SU}(2) \times \textrm{U}(1)  \rightarrow \textrm{SU}(3) \,$ symmetry}

There is a two-parameter family of $\,\mathcal{N}=1\,$ supersymmetric AdS$_{4}$ solutions that preserves $\,\textrm{U}(1)^2\,$ and is located at
\begin{equation}
\label{VEVs_z_N1}
z_{1,2,3} = c \left( -\chi_{1,2,3} +  i \, \frac{\sqrt{5}}{3} \right)
\hspace{6mm} \textrm{and} \hspace{6mm}
z_{4}=z_{5}=z_{6}=z_{7}= \frac{1}{\sqrt{6}} ( 1+ i \, \sqrt{5}) \ ,
\end{equation}
subject to the constraint
\begin{equation}
\chi_{1}+\chi_{2}+\chi_{3}=0 \ .
\end{equation}
This family of AdS$_{4}$ solutions has a vacuum energy given by
\begin{equation}
V_{0} = -\frac{162}{25 \sqrt{5}} \, \, g^2 \, c^{-1} \ ,
\end{equation}
and a spectrum of normalised scalar masses of the form
\begin{equation}
\label{spectrum_SU3_scalars}
\begin{array}{lll}
m^2 L^2 &=&
 0 \,\,\, ( \times  28)
 \hspace{3mm} , \hspace{3mm}
 4 \pm \sqrt{6} \,\,\, ( \times 2 )
 \hspace{3mm} , \hspace{3mm}
 -2 \,\,\, ( \times 2 ) \ ,  \\[4mm]
& &
 -\frac{14}{9} + 5 \chi _i^2 \pm \frac{1}{3} \sqrt{4 + 45 \chi _i^2}    \,\,\, ( \times 2 )
 \hspace{8mm} i=1,2,3 \ ,
 \\[4mm]
& &
-\frac{14}{9}   +  \frac{5}{4}  \chi _i^2 \pm  \frac{1}{6} \sqrt{16 + 45 \chi _i^2} \,\,\, ( \times 2 )
 \hspace{8mm} i=1,2,3 \ ,
 \\[4mm]
& &
\frac{7}{9} +  \frac{5}{4} \chi _i^2  \,\,\, ( \times 2 )
 \hspace{8mm} i=1,2,3 \ ,
\\[4mm]
& &
-2 + \frac{5}{4} \left(\chi_i- \chi_j\right){}^2 \,\,\, ( \times 2 )
\hspace{8mm} i <  j  \ ,  \\[4mm]

\end{array}
\end{equation}
where $\,L^2=-3/V_{0}\,$ is the AdS$_{4}$ radius. The computation of the vector masses yields
\begin{equation}
\label{spectrum_SU3_vectors}
\begin{array}{lll}
m^2 L^2 &=&
0 \,\,\, ( \times 2 )
\hspace{2mm} , \hspace{2mm}
6 \,\,\, ( \times 1 )
\hspace{2mm} , \hspace{2mm}
2 \,\,\, ( \times 1 ) \ ,
\\[4mm]
& &
\frac{16}{9}  +  \frac{5}{4} \chi _i^2 \pm \frac{1}{6} \sqrt{64 + 45 \chi _i^2}   \,\,\, ( \times 2 )
\hspace{8mm} i=1,2,3  \ ,
\\[4mm]
& &
\frac{25}{9} + \frac{5 \chi _i^2}{4}  \,\,\, ( \times 2 )
\hspace{8mm} i=1,2,3 \ , \\[4mm]
& &

\frac{5}{4} \left(\chi _i-\chi _j \right){}^2 \,\,\, ( \times 2 )
\hspace{8mm} i <  j  \ .
\end{array}
\end{equation}
Note that a generic solution in this family preserves $\,\textrm{U}(1)^2\,$ as only two vectors are generically massless. Therefore, out of the $\,28\,$ massless scalars in \eqref{spectrum_SU3_scalars}, only $\,2\,$ of them correspond to physical directions in the potential. The residual symmetry gets enhanced to $\,\textrm{SU}(2) \times \textrm{U}(1)\,$ when imposing a pairwise identification between the axions $\,\chi_{1,2,3}\,$ so that a total of four vectors become massless. Finally there is a symmetry enhancement to $\,\textrm{SU}(3)\,$ when setting $\,\chi_{1,2,3}=0\,$ so that a total of eight vectors become massless. The $\textrm{SU}(3)$ symmetric solution was recently uplifted to a ten-dimensional family of type IIB S-fold backgrounds in \cite{Guarino:2019oct}.


\subsubsection{$\mathcal{N}=2\,$ vacua with $\,\textrm{U}(1)^2 \rightarrow \textrm{SU}(2) \times \textrm{U}(1) \,$ symmetry}

\label{sec:N=2_AdS4_vacua}

There is a one-parameter family of $\,\mathcal{N}=2\,$ supersymmetric AdS$_{4}$ solutions that preserves $\,\textrm{U}(1)^2\,$ and is located at
\begin{equation}
\label{VEVs_z_N2}
z_{1}=-\bar{z}_{3}= c \left( -\chi \, + \,  i \, \frac{1}{\sqrt{2}} \right)
\hspace{3mm} \textrm{,} \hspace{3mm}
z_{2} = i \, c
\hspace{3mm} \textrm{,} \hspace{3mm}
z_{4}=z_{6}= i
\hspace{2mm} \textrm{ and } \hspace{2mm}
z_{5}=z_{7} =  \frac{1}{\sqrt{2}} (1 \, + \, i ) \ .
\end{equation}
This family of AdS$_{4}$ solutions has a vacuum energy given by
\begin{equation}
V_{0} = -3 \,\, g^2 \, c^{-1}\ ,
\end{equation}
and a spectrum of normalised scalar masses of the form
\begin{equation}
\label{spectrum_SU2xU1_scalars}
\begin{array}{lll}
m^2 L^2 &=&
0 \,\,\, ( \times 30 )
\hspace{3mm} , \hspace{3mm}
3 \pm \sqrt{17} \,\,\, ( \times 2 )
\hspace{3mm} , \hspace{3mm}
-2 \,\,\, ( \times 4 )
\hspace{3mm} , \hspace{3mm}
2 \,\,\, ( \times 6 )
\hspace{3mm} , \hspace{3mm}
-2 + 4 \chi^2 \,\,\, ( \times 6 )  \\[2mm]
& & -1 + 4 \chi^2 \pm \sqrt{16 \chi^2+1} \,\,\, ( \times 2 )
\hspace{3mm} , \hspace{3mm}
\chi^2 \pm \sqrt{\chi^2+2} \,\,\, ( \times 8 ) \ ,
\end{array}
\end{equation}
where $\,L^2=-3/V_{0}\,$ is the AdS$_{4}$ radius. The computation of the vector masses yields
\begin{equation}
\label{spectrum_SU2xU1_vectors}
\begin{array}{lll}
m^2 L^2 &=&
0 \,\,\, ( \times  2 )
\hspace{5mm} , \hspace{5mm}
6 \,\,\, ( \times  2 )
\hspace{5mm} , \hspace{5mm}
4 \,\,\, ( \times  2 )
\hspace{5mm} , \hspace{5mm}
2 \, ( \times  4 ) \ , \\[2mm]
& & 4 \chi^2 \,\,\, ( \times  2 )
\hspace{5mm} , \hspace{5mm}
2 + \chi^2 \pm \sqrt{\chi^2+2} \,\,\, ( \times  8 ) \ .
\end{array}
\end{equation}
Note that a generic solution in this family preserves $\,\textrm{U}(1)^2\,$ as only two vectors are generically massless. Therefore, out of the $\,30\,$ massless scalars in \eqref{spectrum_SU2xU1_scalars}, only $\,4\,$ of them correspond to physical directions in the scalar potential. However, the residual symmetry gets enhanced to $\,\textrm{SU}(2) \times \textrm{U}(1)\,$ when $\,\chi=0\,$ and two additional vectors become massless. This special AdS$_{4}$ vacuum will be uplifted to a ten-dimensional family of type IIB S-fold backgrounds in Section~\ref{sec:10D}.


\subsubsection{$\mathcal{N}=4\,$ vacuum with $\,\textrm{SO}(4)\,$ symmetry}

There is an $\,\mathcal{N}=4\,$ supersymmetric AdS$_{4}$ solution that preserves $\,\textrm{SO}(4)\,$ and is located at
\begin{equation}
\label{VEVs_z_N4}
z_{1}=z_{2}=z_{3} =  i \, c
\hspace{8mm} \textrm{and} \hspace{8mm}
z_{4}=z_{5}=z_{6}=-\bar{z}_{7}=  \frac{1}{\sqrt{2}} ( 1+ i ) \ .
\end{equation}
This AdS$_{4}$ solution has a vacuum energy given by
\begin{equation}
V_{0} = -3 \,\, g^2 \, c^{-1} \ ,
\end{equation}
as for the previous solution, and a spectrum of normalised scalar masses of the form
\begin{equation}
\label{spectrum_SO4_scalars}
m^2 L^2 \,\,=\,\,
0 \, ( \times 48)
\hspace{5mm} , \hspace{5mm}
10 \, ( \times 1)
\hspace{5mm} , \hspace{5mm}
 4 \, ( \times 10)
 \hspace{5mm} , \hspace{5mm}
-2 \, ( \times 11)  \ ,
\end{equation}
where $\,L^2=-3/V_{0}\,$ is the AdS$_{4}$ radius.  The computation of the vector masses yields
\begin{equation}
\label{spectrum_SO4_vectors}
m^2 L^2 \,\, =  \,\,
0 \, ( \times 6 )
 \hspace{5mm} , \hspace{5mm}
6 \, ( \times 7 )
\hspace{5mm} , \hspace{5mm}
2 \, ( \times 15 )
 \ ,
\end{equation}
thus reflecting the $\,\textrm{SO}(4)\,$ residual symmetry at the AdS$_{4}$ solution. Therefore, out of the $\,48\,$ massless scalars in \eqref{spectrum_SO4_scalars}, only $\,26\,$ of them correspond to physical directions in the scalar potential. This $\,\mathcal{N}=4\,$ solution was first reported in \cite{Gallerati:2014xra}, and then uplifted to a ten-dimensional family of type IIB S-fold backgrounds in \cite{Inverso:2016eet}.

\section{S-folds with $\,\mathcal{N}=2\,$ supersymmetry}
\label{sec:10D}

From this moment on we will set
\begin{equation}
g=c=1 \ ,
\end{equation}
without loss of generality. From (\ref{VEVs_z_N0}), (\ref{VEVs_z_N1}), (\ref{VEVs_z_N2}) and (\ref{VEVs_z_N4}) it becomes clear that varying $\,c\,$ amounts to a rescaling of the vacuum expection values of $\,z_{1,2,3} \propto c \,$ at the AdS$_{4}$ vacua. After $\,c\,$ has been set to unity,  varying $\,g\,$ simply corresponds to a rescaling of the vacuum energy $\,V_{0} \propto g^{2} \, c^{-1}\,$ and thus to a redefinition of the AdS$_{4}$ radius $\,L^2=-3/V_{0}\,$. Let us emphasise again that all the AdS$_4$ vacua in Section~\ref{sec:AdS4_vacua} are genuinely dyonic as they do not survive the limit $\,c \rightarrow 0\,$ to implement a purely electric gauging. In this limit one has that $\,\textrm{Im} (z_{1,2,3}) \rightarrow 0\,$ or, by virtue of (\ref{Phi_def_7}), a runaway behaviour towards the boundary of moduli space $\,\varphi_{1,2,3} \rightarrow \infty\,$.

Going back to the goal of this section, the $\,\mathcal{N}=2\,$ family of solutions in Section~\ref{sec:N=2_AdS4_vacua} is new and preserves a $\,\textrm{U}(1)^2\,$ symmetry. It is a one-parameter family of AdS$_{4}$ vacua and, in the special case of the parameter vanishing $\,\chi=0\,$, there is an enhancement of symmetry to $\,\textrm{SU}(2) \times \textrm{U}(1)\,$. Following \cite{Inverso:2016eet}, and implementing a generalised S--S ansatz in $\textrm{E}_{7(7)}$-EFT \cite{Hohm:2013uia}, we will uplift such an $\,\mathcal{N}=2\,\, \& \,\, \textrm{SU}(2) \times \textrm{U}(1)\,$ symmetric AdS$_{4}$ vacuum to a class of ten-dimensional S-fold backgrounds of type IIB supergravity of the form $\,\textrm{AdS}_{4} \times \textrm{S}^{1} \times \textrm{S}^{5}\,$ with an S-duality  hyperbolic monodromy along $\,\textrm{S}^{1}\,$.

\subsection{Type IIB uplift using $\textrm{E}_{7(7)}$-EFT}
\label{sec:E77_uplift}

Generalised Scherk--Schwarz (S--S) reductions of exceptional field theory (EFT) have proved a very efficient method to perform consistent truncations of eleven-dimensional and type IIB supergravity on spheres and hyperboloids \cite{Hohm:2014qga}. Here we are interested in the uplift of an AdS$_{4}$ vacuum of a four-dimensional gauged maximal supergravity, which thus selects the $\textrm{E}_{7(7)}$-EFT of \cite{Hohm:2013uia} as the natural framework to carry out this mission.

The $\textrm{E}_{7(7)}$-EFT lives in an extended space-time that consists of an external four-dimensional space with coordinates $\,x^{\mu}\,$ ($\,\mu=0, \ldots, 3\,$) and a 56-dimensional generalised internal space with coordinates $\,Y^{M}\,$ ($\,M=1, \ldots, 56\,$) in the fundamental representation $\textbf{56}$ of $\,\textrm{E}_{7(7)}\,$, subject to the action of the $\,\textrm{E}_{7(7)}$-covariant generalised diffeomorphisms. In order to uplift an AdS$_{4}$ vacuum amongst those in Section~\ref{sec:AdS4_vacua} to a ten-dimensional background of type IIB supergravity, the relevant field content of $\textrm{E}_{7(7)}$-EFT reduces to the external metric $\,g_{\mu\nu}(x,Y)\,$ and the internal generalised metric $\,\mathcal{M}_{MN}(x,Y)\,$ (vector and tensor fields are consistently set to zero). These are connected with the metric $\,g_{\mu\nu}(x)\,$ and the scalar fields $\,M_{MN}(x)\,$ of the four-dimensional maximal supergravity in (\ref{Lagrangian}) via a generalised S--S ansatz \cite{Hohm:2014qga}
\begin{equation}
\label{S--S_ansatz}
\begin{array}{rll}
g_{\mu\nu}(x,Y) &=& \rho^{-2}(Y) g_{\mu\nu}(x) \ , \\[2mm]
\mathcal{M}_{MN} (x,Y) &=& U_M{}^K(Y) \, U_N{}^L(Y) \, M_{KL}(x) \ .
\end{array}
\end{equation}
The entire dependence on the $\,Y^{M}\,$ coordinates is then encoded in a twist matrix $\,U_M{}^K(Y)\,$ and a scaling function $\,\rho(Y)\,$ satisfying
\begin{equation}
\label{S--S_twist}
\begin{array}{rcl}
\left. (U^{-1})_M{}^P \, (U^{-1})_N{}^Q \, \partial_P U_Q{}^K \right|_{\textbf{912}} &=& \tfrac{1}{7} \, \rho \, X_{MN}{}^{K} \ ,  \\[3mm]
\partial_N (U^{-1})_M{}^N - 3 \, \rho^{-1} \, \partial_N \rho \,  (U^{-1})_M{}^N &=& 2 \, \rho \, \vartheta_M \ ,
\end{array}
\end{equation}
where $\,X_{MN}{}^{K}\,$ is the embedding tensor specifying the gauging in the four-dimensional supergravity, $\,\vartheta_{M}\,$ is a constant scaling tensor and $\,|_{\textbf{912}}\,$ denotes projection onto the $\,\textbf{912}\,$ irreducible representation of $\,\textrm{E}_{7(7)}\,$ where the embedding tensor lives.

For the dyonic gauging of $\,\textrm{G} \subset \textrm{SL}(8)\,$ in (\ref{G-group}) the non-vanishing components of the embedding tensor were given in (\ref{X_tensor}) and the tensor $\,\vartheta_{M}\,$ vanishes identically. The generalised S--S ansatz depends on six physical coordinates $\,(y^{i} \, , \, \tilde{y}) \in Y^{M}\,$: five of them are electric $\,y^{i}\,$ $\,(i=2, \ldots , 6)\,$ and one is magnetic $\,\tilde{y}\,$. Considering the electric-magnetic splitting of generalised coordinates $\,Y^{M}=(Y^{AB} \,,\, Y_{AB})\,$ under $\,\textrm{SL}(8) \subset \textrm{E}_{7(7)}\,$, one has
\begin{equation}
y^{i} = Y^{i7} \in Y^{AB}
\hspace{10mm} \textrm{ and } \hspace{10mm}
\tilde{y}=Y_{18} \in Y_{AB} \ .
\end{equation}
In terms of the physical coordinates $\,(y^{i} \,,\, \tilde{y})\,$ the scaling function $\,\rho\,$ in (\ref{S--S_ansatz})-(\ref{S--S_twist}) reads
\begin{equation}
\label{rho_func}
\rho(y^{i} , \tilde{y}) = \hat{\rho}(y^{i})  \, \mathring{\rho}(\tilde{y}) \ ,
\end{equation}
where the two factors in (\ref{rho_func}) are given by
\begin{equation}
\label{rho_hat/cir_func}
\hat{\rho}^{4} =1- \, |\vec{y}|^2
\hspace{10mm} \textrm{ and } \hspace{10mm}
\mathring{\rho}^{4} =1+ \, \tilde{y}^2  \, ,
\end{equation}
and $\,\vec{y}\equiv (y^i)\,$. On the other hand, the generalised twist matrix $\,(U^{-1})_M{}^N \,$ in (\ref{S--S_ansatz})-(\ref{S--S_twist}) is $\,\textrm{SL}(8)$-valued and possesses a block diagonal structure
\begin{equation}
(U^{-1})_M{}^N \,=\, \begin{pmatrix}
(U^{-1})_{\left[AB\right]}{}^{\left[CD\right]} & 0 \\[2mm]
0 & (U^{-1})^{\left[AB\right]}{}_{\left[CD\right]} = U_{\left[CD\right]}{}^{\left[AB\right]}
\end{pmatrix} \ ,
\end{equation}
with components
\begin{equation}
(U^{-1})_{[AB]}{}^{[CD]} \,=\, 2\, (U^{-1})_{[A}{}^{[C} \, (U^{-1})_{B]}{}^{D]}  \ ,
\end{equation}
and
\begin{equation}
\label{U_inv_twist}
(U^{-1})_{A}{}^{B} \,=\, \left(\frac{\mathring{\rho}}{\hat{\rho}}\right)^\frac{1}{2} \,
\left(
\begin{array}{cccc}
1 & 0 & 0 &  \mathring{\rho}^{-2}  \, \tilde{y} \\[2mm]
0 & \delta^{ij} + \hat{K} \, y^{i} \, y^{j} &  \hat{\rho}^2 y^{i} \,  & 0 \\[2mm]
0 &   \hat{\rho}^2 y^{j} \, \hat{K} \,  &  \hat{\rho}^4  & 0 \\[2mm]
\mathring{\rho}^{-2}  \, \tilde{y}  & 0 & 0 & \mathring{\rho}^{-4} (1+\tilde{y}^2)
\end{array}
\right) \ .
\end{equation}
The twist matrix in (\ref{U_inv_twist}) also depends on a function $\,\hat{K}(y^{i})\,$ which is given in this case by a hypergeometric function \cite{Inverso:2016eet}
\begin{equation}
\hat{K} = -\phantom{}_2F_1\left(\, 1\,,\,2\,,\,\tfrac{1}{2} \, ; \, \,1-|\vec{y}|^2 \, \right) \ .
\end{equation}

Using the dictionary between the fields of type IIB supergravity and those of $\textrm{E}_{7(7)}$-EFT  \cite{Baguet:2015xha,Baguet:2015sma}, together with the S--S ansatz (\ref{S--S_ansatz}) involving generalised twist parameters (\ref{rho_func})-(\ref{U_inv_twist}), one arrives at the final uplift formulae
\begin{equation}
\label{uplift_internal_formulae}
\begin{array}{rll}
G^{mn} &=& G^{\frac{1}{2}} \, \mathcal{M}^{mn} \ , \\[2mm]
\mathbb{B}_{mn}{}^\alpha &=& G^{\frac{1}{2}} \, G_{mp} \, \epsilon^{\alpha\beta} \, \mathcal{M}^{p}{}_{n\beta} \ , \\[2mm]
C_{klmn} - \frac{3}{2} \, \epsilon_{\alpha\beta} \, \mathbb{B}_{k\left[l\right.}{}^\alpha \,  \mathbb{B}_{\left.mn\right]}{}^\beta &=& - \frac{1}{2} \, G^{\frac{1}{2}} \,  G_{k p} \, \mathcal{M}^p{}_{lmn} \ ,\\[2mm]
m_{\alpha\beta} &=& \frac{1}{6} \, G \, \left( \, \mathcal{M}^{mn}\, \mathcal{M}_{m\alpha\,n\beta} + \mathcal{M}^m{}_{k\alpha} \, \mathcal{M}^k{}_{m\beta} \, \right) \ ,
\end{array}
\end{equation}
for the purely internal components of the type IIB fields: (inverse) metric $\,G^{mn}\,$, two-form potentials $\,\mathbb{B}^{\alpha}=(B_{2}\,,\,C_{2})\,$ with $\,\alpha = 1, 2\,$, four-form potential $\,C_{4}\,$ and axion-dilaton $\,m_{\alpha \beta}\,$. The various blocks $\,\mathcal{M}^{mn}\,$, $\, \mathcal{M}^{p}{}_{n\beta} \,$, $\,\mathcal{M}^p{}_{lmn}\,$ and $\,\mathcal{M}_{m\alpha\,n\beta}\,$ entering the r.h.s of (\ref{uplift_internal_formulae}) can be extracted from the internal generalised metric $\,\mathcal{M}_{MN} (x,y^{i},\tilde{y}) \,$ by performing the group-theoretical decomposition that is relevant for the embedding of type IIB supergravity into $\textrm{E}_{7(7)}$-EFT:
\begin{equation}
\label{GL(6)_branching}
\begin{array}{ccc}
\textrm{E}_{7(7)} & \supset & \textrm{GL}(6) \,\times \, \textrm{SL}(2)_{\textrm{IIB}}\,  \times \, \mathbb{R}^{+} \\[3mm]
\textbf{56} & \rightarrow & (\textbf{6},\textbf{1})_{+2} \, + \, (\textbf{6'},\textbf{2})_{+1}  \, + \, (\textbf{20},\textbf{1})_{0} \, + \, (\textbf{6},\textbf{2})_{-1} \, + \, (\textbf{6'},\textbf{1})_{-2} \\[3mm]
Y^{M} & \rightarrow & y^{m} \,\, + \,\, y_{m\alpha}  \,\, + \,\, y^{mnp}  \,\,  + \,\, y^{m\alpha} \,\, +  y_{m} \,\,
\end{array}
\end{equation}
The physical coordinates are identified as $\,y^{m}=(y^{i},\tilde{y})\,$, with $\,m=(i,7)\,$ and $\,i=2,\ldots,6\,$, which implies a further group-theoretical branching $\,\textrm{GL}(6) \rightarrow \textrm{GL}(1) \times \textrm{GL}(5)\,$ compatible with the $\,\mathbb{R} \,(\textrm{or } \textrm{S}^1) \times \textrm{S}^5\,$ factorisation of the geometry we are behind of. The various mappings between coordinates discussed above are summarised as
\begin{equation}
\begin{array}{c|c|c|c|c}
 \,\,\,\,\,y^{m} \,\,\,\,& \,\,\,\,y_{m\alpha}  \,\,\,\, &\,\,\,\,\,\, y^{mnp} \,\,\,\, & \,\,\,\,\,\, y^{m\alpha} \,\,\,\, &  y_{m} \\[2mm]
\,\,y^{i} \,\,\,\,\,\,\,\,\,\,  y^{7} \,\,\,\,\,\,\,\, & \,\,y_{i\alpha} \,\,\,\,\,\,\,\,\,\,\,  y_{7\alpha} &  \,\,\,\,\,\,\,\,\,\,y^{ijk}  \,\,\,\,\,\,\,\,\,\,\,\,\,\,\,\,\,\,\, y^{ij7} \,\,\,\, & \,\,\,\, y^{i\alpha} \,\,\,\,\,\,\,\,\,   y^{7 \alpha} \,\,\,\, & y_{i} \,\,\,\,\,\,\,\,  y_{7} \\[2mm]
Y^{i7} \,\,\,\,\,\, Y_{18} \,\,\,\,\,\,\,\, & \,\,\,\,\,\,\, Y_{i \alpha} \,\,\,\,\, \epsilon_{\alpha \beta} \, Y^{\beta 7} &\,\,\,\,   \epsilon^{ijk j'k'}\, Y_{j'k'} \,\,\,\,\,\,\,  Y^{ij} \,\,\,\, &\,\,\,\,  Y^{i\alpha} \,\,\,\,\,  \epsilon^{\alpha \beta} \, Y_{\beta 7} \,\,\,\, & \,\,\,Y_{i7}  \,\,\,\,\,\,\,\, Y^{18}
\end{array}
\end{equation}
We refer the reader to the original works \cite{Baguet:2015xha,Baguet:2015sma} (and also \cite{Inverso:2016eet,Guarino:2019oct}) for more details on the generalised S--S reductions of $\textrm{E}_{7(7)}$-EFT and their connection with the gauged maximal supergravities.

We now move to the uplift the AdS$_{4}$ vacuum with $\,\mathcal{N}=2 \,\, \& \,\, \textrm{SU}(2)\times\textrm{U}(1)\,$ symmetry discussed in Section~\ref{sec:N=2_AdS4_vacua} to a ten-dimensional background of type IIB supergravity using (\ref{uplift_internal_formulae}). We have explicitly verified that the ten-dimensional equations of motion and Bianchi identities of type IIB supergravity are satisfied\footnote{We adopt the type IIB conventions in the appendix~B of \cite{Guarino:2019oct}.}.

\subsubsection*{Ten-dimensional metric}

We adopt the conventions of \cite{Guarino:2019oct} to describe the geometry of the round five-sphere $\,\textrm{S}^5\,$. Using coordinates $\,y^{i}\,$ ($\,i=2,\ldots,6\,$) to parameterise $\,\textrm{S}^{5}\,$, the metric and its inverse are given by
\begin{equation}
\hat{G}_{ij} = \delta_{ij} + \dfrac{\delta_{ik} \, \delta_{jl} \, y^{k} y^{l}}{1-y^{m} \, \delta_{mn} \, y^{n}}
\hspace{10mm} \textrm{ and } \hspace{10mm}
\hat{G}^{ij} = \delta^{ij} - y^{i} y^{j} \ .
\end{equation}
However it will also be convenient to introduce a set of embedding coordinates $\,\mathcal{Y}_{\underline{m}}\,$ on $\,\mathbb{R}^{6}\,$ ($\,\underline{m}=2,\ldots,7\,$) of the form
\begin{equation}
\mathcal{Y}_{\underline{m}}=\left\lbrace  y^{i} \, , \,  \mathcal{Y}_{7} \equiv \big(1- |\vec{y}|^2 \big)^{\frac{1}{2}} \right\rbrace
\hspace{8mm} \textrm{ with } \hspace{8mm}
\delta^{\underline{mn}} \, \mathcal{Y}_{\underline{m}} \, \mathcal{Y}_{\underline{n}} = 1 \ ,
\end{equation}
so that the Killing vectors on $\,\textrm{S}^5\,$ are constructed as
\begin{equation}
\mathcal{K}_{\underline{mn}}{}^{i} \equiv \hat{G}^{ij} \, \partial_{j} \, \mathcal{Y}_{[\underline{m}} \, \mathcal{Y}_{\underline{n}]}
=
\delta^{i}_{[\underline{m}} \, \mathcal{Y}_{\underline{n}]}  \ .
\end{equation}

Following the derivation of \cite{Inverso:2016eet}, the internal part of the ten-dimensional metric has components in (\ref{uplift_internal_formulae}) given by
\begin{equation}
\label{int_metric}
\begin{array}{llll}
G^{11} & = &  \Delta \, \mathring{\rho}^4 \, M_{18\,18} \,\, = \,\, 2 \, \Delta \, (1+\tilde{y}^{2}) \ , \\[4mm]
G^{1k} & = &  \Delta \, \mathring{\rho}^2 \, \mathcal{K}_{\underline{ij}}{}^{k} \, M^{\underline{ij}}{}_{18} \,\, = \,\, 0  \ , \\[4mm]
G^{ij} & = & \Delta  \, \mathcal{K}_{\underline{kl}}{}^{i} \, \mathcal{K}_{\underline{mn}}{}^{j} \, M^{\underline{kl} \, \underline{mn} } \,\, = \,\,   \Delta \,  \left(  \hat{G}^{ij} +  L^{ij} \right)  \  ,
\end{array}
\end{equation}
where $\,M^{\underline{ij}}{}_{18}=0\,$ as a consequence of having set $\,\chi = 0\,$ in the $\,\mathcal{N}=2\,$ AdS$_{4}$ vacuum, and where we have defined
\begin{equation}
 L^{ij} = \left(
\begin{array}{ccccc}
 \mathcal{Y}_{\underline{4}}^2 +  \mathcal{Y}_{\underline{5}}^2 +  \mathcal{Y}_{\underline{6}}^2 & - \mathcal{Y}_{\underline{6}}  \, \mathcal{Y}_{\underline{7}} & - \mathcal{Y}_{\underline{2}}  \, \mathcal{Y}_{\underline{4}} & - \mathcal{Y}_{\underline{2}}  \, \mathcal{Y}_{\underline{5}} & - \mathcal{Y}_{\underline{2}}  \, \mathcal{Y}_{\underline{6}} \\[2mm]
 - \mathcal{Y}_{\underline{6}}  \, \mathcal{Y}_{\underline{7}} & \mathcal{Y}_{\underline{4}}^2 +  \mathcal{Y}_{\underline{5}}^2 +  \mathcal{Y}_{\underline{7}}^2 & - \mathcal{Y}_{\underline{3}}  \, \mathcal{Y}_{\underline{4}} & - \mathcal{Y}_{\underline{3}}  \, \mathcal{Y}_{\underline{5}} & \phantom{-} \mathcal{Y}_{\underline{2}}  \, \mathcal{Y}_{\underline{7}} \\[2mm]
 - \mathcal{Y}_{\underline{2}}  \, \mathcal{Y}_{\underline{4}} &  - \mathcal{Y}_{\underline{3}}  \, \mathcal{Y}_{\underline{4}} & 1 - \mathcal{Y}_{\underline{4}}^2 & - \mathcal{Y}_{\underline{4}}  \, \mathcal{Y}_{\underline{5}} & - \mathcal{Y}_{\underline{4}}  \, \mathcal{Y}_{\underline{6}} \\[2mm]
 - \mathcal{Y}_{\underline{2}}  \, \mathcal{Y}_{\underline{5}} &  - \mathcal{Y}_{\underline{3}}  \, \mathcal{Y}_{\underline{5}} &  - \mathcal{Y}_{\underline{4}}  \, \mathcal{Y}_{\underline{5}} & 1 - \mathcal{Y}_{\underline{5}}^2 & - \mathcal{Y}_{\underline{5}}  \, \mathcal{Y}_{\underline{6}} \\[2mm]
- \mathcal{Y}_{\underline{2}}  \, \mathcal{Y}_{\underline{6}} & \phantom{-}  \mathcal{Y}_{\underline{2}}  \, \mathcal{Y}_{\underline{7}} & - \mathcal{Y}_{\underline{4}}  \, \mathcal{Y}_{\underline{6}} & - \mathcal{Y}_{\underline{5}}  \, \mathcal{Y}_{\underline{6}} &  \mathcal{Y}_{\underline{4}}^2 +  \mathcal{Y}_{\underline{5}}^2 +  \mathcal{Y}_{\underline{2}}^2
 \end{array}
 \right) \ .
\end{equation}
The warping factor $\,\Delta\,$ in (\ref{int_metric}) is nowhere vanishing and reads
\begin{equation}
\Delta = (\det G)^{\frac{1}{2}} \, \rho^2 = \frac{1}{\sqrt{2}} \, \left( 1 + \mathcal{Y}_{\underline{4}}^2 + \mathcal{Y}_{\underline{5}}^2 \right)^{-\frac{1}{4}} \ .
\end{equation}

The six-dimensional internal metric becomes more transparent if first introducing a new variable for the magnetic coordinate
\begin{equation}
\tilde{y} = \sinh\eta
\hspace{8mm} \textrm{ with } \hspace{8mm}
\eta \in (-\infty , \infty ) \ ,
\end{equation}
and then a set of angular variables for $\,\textrm{S}^5\,$ of the form
\begin{equation}
\begin{array}{c}
y^2 = \cos\theta  \cos\left(\frac{\beta}{2}\right)   \cos \left(\frac{\alpha +\gamma}{2} \right)
\,\,\,\,\,\, , \,\,\,\,\,\,
y^3 =  \cos\theta   \cos\left(\frac{\beta}{2}\right)     \sin\left(\frac{\alpha +\gamma}{2}\right) \ , \\[4mm]
y^4 = \cos\phi   \sin\theta
\,\,\,\,\,\, , \,\,\,\,\,\,
y^5 = \sin\phi   \sin\theta
\,\,\,\,\,\, , \,\,\,\,\,\,
y^6 = \cos\theta    \sin\left(\frac{\beta}{2}\right)   \cos \left(\frac{\alpha -\gamma}{2}\right) \ ,
\end{array}
\end{equation}
with ranges given by
\begin{equation}
\theta \in [0,\tfrac{\pi}{2}]
\hspace{4mm} , \hspace{4mm}
\phi \in [0,2\pi]
\hspace{4mm} , \hspace{4mm}
\alpha \in [0,2\pi]
\hspace{4mm} , \hspace{4mm}
\beta \in [0,\pi]
\hspace{4mm} , \hspace{4mm}
\gamma \in [0,4\pi] \ .
\end{equation}
In this manner, and upon introducing a set of $\,\textrm{SU}(2)\,$ left-invariant one-forms
\begin{equation}
\begin{array}{llll}
\sigma_{1} &=& \frac{1}{2} \,  (  - \sin\alpha \, d\beta + \cos\alpha \, \sin\beta \, d\gamma  ) & , \\[2mm]
\sigma_{2} &=& \frac{1}{2} \,  ( \cos\alpha \, d\beta + \sin\alpha \, \sin\beta  \,  d\gamma ) & , \\[2mm]
\sigma_{3} &=& \frac{1}{2} \,  ( d\alpha + \cos\beta \, d\gamma ) & ,
\end{array}
\end{equation}
the internal six-dimensional metric takes a simple $\,\mathbb{R} \times \textrm{S}^5\,$ form
\begin{equation}
\label{ds6}
ds_6^2 = \frac{1}{2} \, \Delta^{-1} \left[ d\eta^2 + ds^2_{\textrm{S}^2}  +  \cos^2\theta \, ds^2_{\textrm{S}^3} \right] \ ,
\end{equation}
with a warping factor
\begin{equation}
\label{Delta_func}
\Delta^{-1} = \big( \, 6-2 \cos (2 \theta ) \, \big)^{\frac{1}{4}} \ ,
\end{equation}
and where we have introduced $\,\textrm{S}^{2}\,$ and (squashed) $\,\textrm{S}^{3}\,$ metrics to describe the deformation of the internal $\,\textrm{S}^{5}\,$. These metrics are explicitly given by
\begin{equation}
ds^2_{\textrm{S}^2} = d\theta^2 + \sin^2\theta \, d\phi^2
\hspace{10mm} \textrm{ and } \hspace{10mm}
ds^2_{\textrm{S}^3} =\sigma _2^2  + 8 \, \Delta^{4} \, \left(\sigma _1^2+\sigma _3^2\right) \ .
\end{equation}

Bringing together (\ref{ds6}) and the external AdS$_4$ part of the geometry, one obtains a ten-dimensional metric of the form\footnote{Restoring the explicit dependence of the warping factor (\ref{Delta_func}) on the parameter $\,c\,$ one finds $\,\Delta \propto c \,$, so the (electric) limit $\,c \rightarrow 0\,$ of the metric (\ref{metric_IIB}) becomes pathological. In other words, the ten-dimensional solution is genuinely dyonic, namely, it requires $\,c \neq 0\,$, as for its associated AdS$_{4}$ vacuum in (\ref{VEVs_z_N2}) with $\,\chi=0\,$.}
\begin{equation}
\label{metric_IIB}
\begin{array}{lll}
ds^{2} &=&  \frac{1}{2} \, \Delta^{-1} \left[ ds_{\text{AdS}_4}^2 +  d\eta^2 + ds^2_{\textrm{S}^2}  +  \cos^2\theta \, ds^2_{\textrm{S}^3} \right] \ .
\end{array}
\end{equation}
This metric has an $\,\textrm{SU(2)} \times \textrm{U}(1)_{\phi}  \times \textrm{U}(1)_{\sigma} \,$ symmetry, where $\,\textrm{U}(1)_{\sigma}\,$ acts as a rotation on the $(\sigma_{1},\sigma_{3})$-plane. Finally, our choice of \textit{undeformed} frames for the metric (\ref{metric_IIB}) is
\begin{equation}
\begin{array}{llllll}
ds^2_{\textrm{AdS}_{4}} & : & \hat{e}^{0}=\dfrac{L}{r} \, dr  \,\,\,\, , \,\,\,\,  \hat{e}^{i}=\dfrac{L}{r}\, dx^{i} \,\,\,\,  \hspace{5mm} (i=1,2,3) \hspace{5mm} \textrm{ and } \hspace{5mm} \eta_{ij}=(-1,1,1)\\[2mm]
ds^2_{\mathbb{R}} & : & \hat{e}^{4} = d\eta  \\[2mm]
ds^2_{\textrm{S}^2}  & : & \hat{e}^{5} = d \theta  \,\,\,\, , \,\,\,\, \hat{e}^{6} = \sin\theta \, d\phi  \\[2mm]
ds^2_{\textrm{S}^3} & : & \hat{e}^{7} = \sigma_{1}  \,\,\,\,\,\, , \,\,\,\,\,\,  \hat{e}^{8} =   \sigma_{2}   \,\,\,\,\,\, , \,\,\,\,\,\,  \hat{e}^{9} =   \sigma_{3}  \\[2mm]
\end{array}
\end{equation}
with $\,L^2=-3/V_{0}=1\,$ being the AdS$_{4}$ radius at the four-dimensional $\,\mathcal{N}=2\,\, \& \,\, \textrm{SU}(2) \times \textrm{U}(1)\,$ symmetric AdS$_{4}$ vacuum.

\subsubsection*{$B_{2}$ and $C_{2}$ potentials}

The two-form potentials $\,\mathbb{B}^{\alpha}=(B_{2}\, ,\,C_{2})\,$ in (\ref{uplift_internal_formulae}) transform as a doublet under the global S-duality group $\,\textrm{SL}(2,\mathbb{R})_{\textrm{IIB}}\,$ of type IIB supergravity. An explicit computation along the lines of \cite{Inverso:2016eet} shows that
\begin{equation}
\begin{array}{lll}
\label{B2C2_uplift}
\mathbb{B}_{1j}{}^{\alpha} &=& 0 \ , \\[2mm]
\mathbb{B}_{ij}{}^{\alpha} &=& \Delta \, G_{ik} \, \mathcal{K}_{\underline{kl}}{}^{k} \, \partial_{j}\mathcal{Y}^{\underline{m}} \, \epsilon^{\alpha \beta} \, (A^{-1})^{\gamma}{}_{\beta} \, M^{\underline{kl}}{}_{\underline{m}\gamma} \ ,
\end{array}
\end{equation}
in terms of a local $\,\textrm{SO}(1,1) \subset \textrm{SL}(2,\mathbb{R})_{\textrm{IIB}}\,$ twist matrix
\begin{equation}
\label{Ainv_twist}
A^{\alpha}{}_{\beta} \equiv
\left(
\begin{array}{cc}
\cosh\eta & \sinh\eta \\[2mm]
\sinh\eta & \cosh\eta
\end{array}
\right)
\hspace{5mm} , \hspace{5mm}
(A^{-1})^{\gamma}{}_{\beta} \equiv
\left(
\begin{array}{cc}
\cosh\eta & -\sinh\eta \\[2mm]
- \sinh\eta & \cosh\eta
\end{array}
\right) \ .
\end{equation}
This matrix encodes the dependence of the two-form potentials on the direction $\,\eta\,$. Using the scalar block $\,M^{\underline{kl}}{}_{\underline{m}\gamma}\,$ at the $\,\mathcal{N}=2\,$ AdS$_{4}$ vacuum under consideration, and using differential form notation, one finds
\begin{equation}
\label{BC_IIB}
\mathbb{B}^{\alpha}  =  A^{\alpha}{}_{\beta} \,\,  \mathfrak{b}^{\beta}\ ,
\end{equation}
with
\begin{equation}
\label{b_frak_IIB}
\begin{array}{llrr}
\mathfrak{b}^{1} & = & \frac{1}{\sqrt{2}} \, \cos\theta \,
\left[ \, \left( \, \cos\phi \, d\theta + \tfrac{1}{2} \, \sin(2 \theta ) \, d(\cos \phi ) \, \right)  \wedge \sigma_2
+ \cos\phi \,\, \dfrac{4 \, \sin (2\theta )}{6-2\cos (2 \theta )} \,\, \sigma_1\wedge \sigma_3  \, \right] & , \\[6mm]
\mathfrak{b}^{2} & = & -\frac{1}{\sqrt{2}} \, \cos\theta \,
\left[ \, \left( \, \sin\phi \, d\theta +  \tfrac{1}{2} \, \sin(2 \theta ) \, d(\sin \phi ) \, \right)  \wedge \sigma_2
+  \sin\phi \,\, \dfrac{4 \, \sin (2\theta )}{6-2\cos (2 \theta )} \,\, \sigma_1\wedge \sigma_3  \, \right]  & .
\end{array}
\end{equation}
The two-form potentials in (\ref{b_frak_IIB}) preserve $\,\textrm{SU(2)} \times \textrm{U}(1)_{\sigma} \,$ but break the $\textrm{U}(1)_{\phi}$ factor due to the explicit dependence on the coordinate $\,\phi\,$.

\subsubsection*{$C_4$ potential}

The internal component of the four-form potential $\,C_{4}\,$ can be explicitly obtained from the third uplift formula in (\ref{uplift_internal_formulae}). Computing the associated (purely internal) five-form field strength, and imposing ten-dimensional self-duality, one gets
\begin{equation}
\label{F5_IIB}
\begin{array}{llll}
\widetilde{F}_5 &=& d C_4 - \tfrac{1}{2}  \, \epsilon_{\alpha\beta}  \, \mathbb{B}^\alpha \wedge \mathbb{H}^\beta & \\[4mm]
&=& (1+\star)  \left[\,  6 \, \sqrt{2} \, \Delta^{5/2} \, \text{vol}_{\textrm{M}_5}  \right. \\[4mm]
&& \left.  \hspace{13mm} - 4 \, \Delta^{4} \, \sin\theta \, \cos^3\theta  \, d \eta \wedge \big(\cos(2\phi) \, d \theta -\tfrac{1}{2} \,   \sin(2\theta) \, \sin(2\phi) \, d\phi  \, \big) \wedge \sigma_1\wedge \sigma_2\wedge \sigma_3 \right] \  ,
\end{array}
\end{equation}
where
\begin{equation}
\label{Vol_M5}
\text{vol}_{\textrm{M}_5}  = \sqrt{2} \, \Delta^{3/2} \,  \sin\theta \, \cos^3\theta \,\,\, d\theta \wedge d\phi \wedge \sigma_1\wedge \sigma_2\wedge \sigma_3 \ ,
\end{equation}
denotes the volume of the deformed five-sphere. Note that $\textrm{U}(1)_{\phi}$ is also broken by $\,\widetilde{F}_5\,$ due to its explicit dependence on the coordinate $\,\phi\,$.

\subsubsection*{Axion-dilaton}

The axion-dilaton matrix $\,m_{\alpha \beta}\,$ can be obtained from the last equation in (\ref{uplift_internal_formulae}). Transforming linearly under S-duality, a direct computation shows an explicit dependence of $\,m_{\alpha \beta}\,$ on the $A$-twist in (\ref{Ainv_twist}) of the form
\begin{equation}
\label{mab_IIB}
m_{\alpha \beta}=
\frac{1}{\textrm{Im}\tau} \, \left(
\begin{array}{cc}
|\tau|^2 & - \textrm{Re}\tau  \\[2mm]
- \textrm{Re}\tau  &  1
\end{array}
\right)
= (A^{-t})_{\alpha}{}^{\gamma} \, \mathfrak{m}_{\gamma \delta} \, (A^{-1})^{\delta}{}_{\beta} \ ,
\end{equation}
with $\,\tau=C_{0}+i \, e^{-\Phi}\,$ and
\begin{equation}
\label{mab_frak_IIB}
\mathfrak{m}_{\gamma \delta} = 2 \, \Delta^2  \,
\left(
\begin{array}{cc}
1 + \sin^2\theta \, \cos^2\phi & -\frac{1}{2} \sin^2\theta \sin (2 \phi ) \\[2mm]
 -\frac{1}{2} \sin^2\theta \sin (2 \phi ) & 1+ \sin^2\theta \, \sin^2\phi
\end{array}
\right) \ .
\end{equation}
Again $\textrm{U}(1)_{\phi}$ is broken by the explicit dependence of (\ref{mab_frak_IIB}) on the angle $\,\phi\,$. This concludes the uplift of the AdS$_{4}$ vacuum with $\,\mathcal{N}=2\,$ and $\,\textrm{SU}(2)\times\textrm{U}(1)\,$ symmetry discussed in Section~\ref{sec:N=2_AdS4_vacua} to a ten-dimensional background of type IIB supergravity. It is worth emphasising that, if trivialising the $A$-twist in (\ref{Ainv_twist}), \textit{i.e.} $\,A^{\alpha}{}_{\beta} =\delta^{\alpha}{}_{\beta} \,$, then the ten-dimensional equations of motion of type IIB supergravity are no longer satisfied.

\subsection{S-fold interpretation}

The dependence of the full type IIB solution on the coordinate $\,\eta\,$ along the $\,\mathbb{R}\,$ direction of the geometry (\ref{metric_IIB}) is totally encoded in the local $\,\textrm{SL}(2,\mathbb{R})_{\textrm{IIB}}\,$ $A$-twist in (\ref{Ainv_twist}). This twist matrix is of hyperbolic type and thus induces a non-trivial monodromy
\begin{equation}
\label{S1_monodromy}
\mathfrak{M}_{\textrm{S}^{1}} =
A^{-1}(\eta) \, A(\eta+T)
=
\left(
\begin{array}{rr}
 \cosh T & \sinh T \\
 \sinh T & \cosh T \\
\end{array}
\right) \ ,
\end{equation}
when forcing the $\,\eta\,$ coordinate to be periodic $\,\eta \rightarrow \eta+T\,$ with period $\,T\,$, namely, when replacing $\,\mathbb{R} \rightarrow \textrm{S}^{1}\,$ in the geometry. Generalising the $A$-twist in (\ref{Ainv_twist}) to a discrete $k$-family ($\,k \in \mathbb{N}\,$ with $\,k\ge 3\,$) of new ones
\begin{equation}
A_{(k)}=A \, g(k)
\hspace{10mm} \textrm{ with } \hspace{10mm}
g(k) =
\left(
\begin{array}{cc}
 \dfrac{(k^2-4)^{\frac{1}{4}}}{\sqrt{2}} & 0 \\[2mm]
 \frac{k}{\sqrt{2} \, (k^2-4)^{\frac{1}{4}}} & \dfrac{\sqrt{2}}{(k^2-4)^{\frac{1}{4}}}
\end{array}
\right) \ ,
\end{equation}
the monodromy (\ref{S1_monodromy}) gets generalised to a $k$-family of $\,\textrm{SL}(2,\mathbb{Z})_{\textrm{IIB}}\,$ hyperbolic monodromies
\begin{equation}
\label{k_monodromy}
\mathfrak{M}(k) =  A^{-1}_{(k)}(\eta)  \,\,\, A_{(k)}\big(\eta+T(k)\big)  =  \left(
\begin{array}{ll}
k  & 1 \\[2mm]
-1  & 0
\end{array}
\right)
\hspace{8mm} \textrm{ , }\hspace{8mm}
k \ge 3 \ ,
\end{equation}
with $\,T(k)=\textrm{ln} (k+\sqrt{k^2-4}) - \textrm{ln}(2)\,$ and $\,\textrm{Tr}\,\mathfrak{M}(k) >2\,$. Therefore, as discussed in \cite{Inverso:2016eet} (see also \cite{Guarino:2019oct}), these backgrounds can be interpreted as locally geometric compactifications on $\,\textrm{S}^{1} \times \textrm{S}^{5}\,$ involving a $k$-family of S-duality monodromies (\ref{k_monodromy}). These monodromies can be written as
\begin{equation}
\mathfrak{M}(k) = - \mathcal{S} \, \mathcal{T}^{k}
\hspace{8mm} \textrm{ with } \hspace{8mm}
\mathcal{S}=\left(
\begin{array}{cc}
0  & -1 \\[2mm]
1  & 0
\end{array}
\right)
\hspace{5mm} \textrm{ and }\hspace{5mm}
\mathcal{T}=\left(
\begin{array}{cc}
1  & 0 \\[2mm]
1  & 1
\end{array}
\right)
\ ,
\end{equation}
and thus define a $k$-family of S-fold backgrounds. Moreover, the argument wielded in \cite{Inverso:2016eet} for the straightforward uplift of the four-dimensional supersymmetries to ten dimensions relied on the monodromy (\ref{S1_monodromy}) being in the hyperbolic conjugacy class of $\,\textrm{SL}(2,\mathbb{R})_{\textrm{IIB}}\,$. This is still our case, so the S-folds presented here preserve $\,\mathcal{N}=2\,$ supersymmetry.

Lastly, various holographic aspects of both $\,\mathcal{N}=4\,$ \cite{Inverso:2016eet} and $\,\mathcal{N}=1\,$ \cite{Guarino:2019oct,Bobev:2019jbi} S-folds with hyperbolic monodromies have respectively been investigated in \cite{Assel:2018vtq,Garozzo:2018kra,Garozzo:2019ejm} and \cite{Bobev:2019jbi} within the context of three-dimensional quiver theories involving $\,\mathcal{N}=4\,$ $\,T(U(N))\,$ theories \cite{Gaiotto:2008ak}, and their potential generalisation to $\,\mathcal{N}=1\,$ SCFT's. It would be interesting to extend these holographic studies to the $\,\mathcal{N}=2\,$ S-folds with hyperbolic monodromies (\ref{k_monodromy}) presented in this work.

\subsection{Connection with Janus-like solutions}

The type IIB solution with $\,{\mathcal{N}=2 \,\, \& \,\, \textrm{SU}(2)\times\textrm{U}(1)}\,$ symmetry we just obtained can be mapped to a new (but equivalent) solution with a linear dilaton profile along the coordinate $\,\eta\,$ upon performing a global $\,\Lambda \in \textrm{SL}(2,\mathbb{R})_{\textrm{IIB}}\,$ transformation, equivalently a change of duality frame, based on the matrix element
\begin{equation}
\label{Lambda_mat}
\Lambda = \frac{1}{\sqrt{2}}
\left(
\begin{matrix}
1 & -1 \\[2mm]
1 & 1
\end{matrix}
\right) \ .
\end{equation}
The composed action of $\,\Lambda \, A^{-1}(\eta)\,$ on (\ref{mab_frak_IIB}) yields a shift of the form $\,\Phi \rightarrow \Phi -2 \eta \,$. Therefore, a degenerate Janus-like behaviour with a linear dilaton $\,\Phi\,$ running from $\,-\infty\,$ to $\,\infty\,$ becomes manifest
\begin{equation}
\label{gs}
g_{s} = e^{\Phi} \propto e^{-2 \eta} \ ,
\end{equation}
giving rise to a varying string coupling $\,g_{s}\,$ that interpolates between the singular values $\,0\,$ and $\,\infty\,$.

Upon performing the $\,\Lambda \in \textrm{SL}(2,\mathbb{R})_{\textrm{IIB}}\,$ transformation (\ref{Lambda_mat}) on the original solution found in Section~\ref{sec:E77_uplift}, a new type IIB background is generated. The metric and self-dual five-form flux are $\,\textrm{SL}(2,\mathbb{R})_{\textrm{IIB}}\,$ singlets and are not affected by the transformation. Therefore, they  take the same form as in (\ref{metric_IIB}) and (\ref{F5_IIB}), namely,
\begin{equation}
\label{new_typeIIB_sol_G&F5}
\begin{array}{lll}
ds^{2} &=&  \frac{1}{2} \, \Delta^{-1} \left[ ds_{\text{AdS}_4}^2 +  d\eta^2 + d\theta^2 + \sin^2\theta \, d\phi^2  +  \cos^2\theta \left(  \sigma _2^2  + 8 \, \Delta^{4} \, \left(\sigma _1^2+\sigma _3^2\right) \right) \right] \ , \\[4mm]
\widetilde{F}_5 &=& 4 \, \Delta^{4} \, \sin\theta \, \cos^3\theta  \,\, (1+\star) \,\, \Big[ \,  3 \,  d\theta \wedge d\phi \wedge \sigma_1\wedge \sigma_2\wedge \sigma_3  \\[2mm]
&&  \hspace{15mm} -   \, d \eta \wedge \big(\cos(2\phi) \, d \theta -\tfrac{1}{2}   \sin(2\theta) \, \sin(2\phi) \, d\phi  \, \big) \wedge \sigma_1\wedge \sigma_2\wedge \sigma_3 \Big] \ .
\end{array}
\end{equation}
The axion-dilaton matrix $\,m_{\alpha \beta}\,$ in (\ref{mab_IIB}) transforms linearly under $\,\textrm{SL}(2,\mathbb{R})_{\textrm{IIB}}\,$. Reading off the new components of $\,\tau\,$ one finds
\begin{equation}
\label{new_typeIIB_sol_RR}
\begin{array}{lll}
\Phi &=& -2 \eta + \log \left[\frac{1}{2} \, \Delta^2 \,  \big( \, 5-\cos(2 \theta) -2 \sin^2\theta \sin(2 \phi) \, \big) \right] \ , \\[4mm]
C_{0} & = & -2 \, e^{2 \eta} \, \dfrac{\cos(2 \phi) \sin^2\theta}{5-\cos(2 \theta) -2 \sin^2\theta \sin(2 \phi) } \ .
\end{array}
\end{equation}
The two-form potentials $\,\mathbb{B}^{\alpha}=(B_{2}\, ,\,C_{2})\,$ in (\ref{BC_IIB})-(\ref{b_frak_IIB}) transform as an $\,\textrm{SL}(2,\mathbb{R})_{\textrm{IIB}}\,$ doublet and take the new form\footnote{The two terms in $\,B_{2}\,$ and $\,C_{2}\,$ which are proportional to $\, \sigma_1\wedge \sigma_3 \,$ can be eliminated by means of a gauge transformation of the form
\begin{equation}
\begin{array}{llll}
B_2 & \rightarrow & B_2 - d \big(  2 \, \sqrt{2} \, \Delta^4 \, e^{-\eta} \, \sin (2 \theta ) \, \cos \theta  \, \cos\psi  \, \sigma_2  \big) & , \\[2mm]
C_2 & \rightarrow & C_2 + d \big(  2 \, \sqrt{2} \, \Delta^4 \, e^{\eta}  \, \sin (2 \theta )  \, \cos \theta \, \sin\psi \, \sigma_2 \big) & ,
\end{array}
\end{equation}
where we have shifted the coordinate $\,\phi \rightarrow \psi + \tfrac{\pi}{4}\,$. However, since these terms are generated by the generalised S--S ansatz discussed in Section~\ref{sec:E77_uplift}, we will retain them here.}
\begin{equation}
\label{new_typeIIB_sol_B2C2}
\begin{array}{lll}
B_{2} & = & e^{-\eta } \,  \left[ \, \frac{1}{2} \,  \cos\theta \, \big(  \,  (\cos\phi + \sin\phi) \, d\theta
+  \frac{1}{2} \, \sin(2\theta) \, (\cos\phi - \sin\phi)   \, d\phi \, \big)  \wedge \sigma_{2} \right.  \\[4mm]
& + & \left. 2 \, \Delta^4 \, \cos\theta \,  \sin(2\theta) \, (\cos\phi + \sin\phi) \, \sigma_{1} \wedge \sigma_{3} \,  \right] \ , \\[6mm]
C_{2} & = &  e^{\eta } \,  \left[ \, \frac{1}{2} \, \cos\theta \, \big(  \,  (\cos\phi - \sin\phi) \, d\theta
-  \frac{1}{2} \, \sin(2\theta) \, (\cos\phi + \sin\phi)   \, d\phi \, \big)  \wedge \sigma_{2} \right. \\[4mm]
& + & \left. 2 \, \Delta^4  \, \cos\theta \,  \sin(2\theta) \, (\cos\phi - \sin\phi) \, \sigma_{1} \wedge \sigma_{3} \, \right]  \ .
\end{array}
\end{equation}
The nowhere vanishing warping factor still reads
\begin{equation}
\Delta^{-4} = 6-2 \cos (2 \theta ) \ .
\end{equation}

In the asymptotic region at $\,\eta \rightarrow -\infty\,$ one has that $\,g_{s}\,$ in (\ref{gs}) diverges (strong coupling) and $\,B_{2}\,$ dominates over other gauge potentials, \textit{e.g.}, $\,C_{0} \rightarrow 0\,$ and $\,C_{2} \rightarrow 0\,$. On the contrary, in the asymptotic region at $\,\eta \rightarrow \infty\,$, the solution becomes dominated by $\,C_{0}\,$ and $\,C_{2}\,$ whereas $\,g_{s}\rightarrow 0\,$ (weak coupling) and $\,B_{2}\rightarrow 0\,$. At intermediate values of the coordinate $\,\eta\,$ one has an interpolating behaviour between these two regimes. Finally, it is also worth noticing that, unlike for the $\,\mathcal{N}=4\,$ \cite{Inverso:2016eet} and $\,\mathcal{N}=1\,$ \cite{Guarino:2019oct} S-folds, there is no $\,\textrm{SL}(2,\mathbb{R})_{\textrm{IIB}}\,$ frame in which the axion $\,C_{0}\,$ (and thus the dual $\theta$-angle) vanishes identically or becomes independent of the coordinate $\,\eta\,$.

\section{Conclusions}
\label{Sec:conclusions}

In this work we have extended the study of AdS$_{4}$ vacua in \cite{DallAgata:2011aa,Gallerati:2014xra,Guarino:2019oct} for the dyonically-gauged $\,[\textrm{SO}(1,1) \times \textrm{SO}(6)] \ltimes \mathbb{R}^{12}\,$ maximal supergravity and found multi-parametric families of new AdS$_{4}$ vacua. Within one such families, all the solutions preserve the same amount of supersymmetry but, importantly, residual symmetry enhancements occur at particular values of the parameters. The previously known $\,\mathcal{N}=0 \,\, \& \,\,\textrm{SO}(6)\,$ \cite{DallAgata:2011aa}, $\,\mathcal{N}=1 \,\, \& \,\,\textrm{SU}(3)\,$ \cite{Guarino:2019oct} and $\,\mathcal{N}=4 \,\, \& \,\,\textrm{SO}(4)\,$ \cite{Gallerati:2014xra} AdS$_{4}$ vacua are shown to correspond to the points of largest symmetry enhancement within their respective families. This is in line with the analysis of (global) symmetry breaking patterns of three-dimensional interface SYM theories presented in \cite{DHoker:2006qeo}.

In the second part of the paper we focused on the new family of $\,\mathcal{N}=2\,$ supersymmetric AdS$_{4}$ vacua and, more concretely, on the vacuum within this family featuring the largest possible residual symmetry, which turns to be $\,\textrm{SU}(2) \times \textrm{U}(1)\,$. By implementing a generalised S--S ansatz in $\,\textrm{E}_{7(7)}$-EFT, we uplifted the AdS$_{4}$ vacuum to a new family of $\,\textrm{AdS}_{4} \times \textrm{S}^{1} \times \textrm{S}^{5}\,$ S-folds of type IIB supergravity with hyperbolic monodromies $\,\mathfrak{M}(k) =-\mathcal{S}\,\mathcal{T}^{k}\,$ (with $\,k \ge 3\,$) along $\,\textrm{S}^1\,$. The residual $\,\textrm{SU}(2) \times \textrm{U}(1)\,$ symmetry and $\,\mathcal{N}=2\,$ supersymmetry of the AdS$_{4}$ vacuum are realised on the S-folds: the internal $\,\textrm{S}^5\,$ is deformed into a product of $\,\textrm{S}^2\,$ and (squashed) $\,\textrm{S}^3\,$ with $\,\textrm{SU}(2) \times \textrm{U}(1)_{\sigma} \times \textrm{U}(1)_{\phi}\,$ isometries and a warping factor, whereas the background fluxes break the $\,\textrm{U}(1)_{\phi}\,$ factor explicitly by introducing a dependence on the coordinate $\,\phi\,$. In many aspects, the realisation of symmetries is much alike the $\,\textrm{AdS}_{5} \times \textrm{S}^{5} \,$ background by Pilch and Warner \cite{Pilch:2000ej} that uplifts the $\,\mathcal{N}=2\,$ and $\,\textrm{SU}(2) \times \textrm{U}(1)\,$ symmetric $\textrm{AdS}_{5}$ vacuum of the five-dimensional SO(6) maximal supergravity presented in \cite{Khavaev:1998fb}.

Finally it would be interesting to investigate the brane setups underlying the families of \mbox{S-folds} presented here (and in \cite{Guarino:2019oct}), especially due to the non-trivial $\,\textrm{SL}(2,\mathbb{Z})_{\textrm{IIB}}\,$ hyperbolic monodromies $\,\mathfrak{M}(k)= - \mathcal{S} \, \mathcal{T}^{k}\,$. It would also be interesting to investigate holographic aspects of such $\,\mathcal{N}=2\,$ and $\,\mathcal{N}=1\,$ S-folds (in the spirit of the $J$-fold CFT's of \cite{Assel:2018vtq,Garozzo:2018kra,Garozzo:2019ejm}\cite{Bobev:2019jbi} with $\,J= - \mathcal{S} \, \mathcal{T}^{k}\,$), as well as to study holographic RG flows by explicitly constructing domain-wall solutions interpolating between the various families of AdS$_{4}$ vacua presented in this work. Lastly, since the S-folds here and in \cite{Guarino:2019oct} display $\,\textrm{SU}(2)\,$ isometries in the internal geometry, it would also be interesting to apply non-abelian T-duality in order to generate new analytic type IIA backgrounds. We plan to address these and related issues in the future.

\section*{Acknowledgements}

We are grateful to Nikolay Bobev, Gianluca Inverso, Yolanda Lozano and Henning Samtleben for interesting conversations. The work of AG is supported by the Spanish government grant PGC2018-096894-B-100 and by the Principado de Asturias through the grant FC-GRUPIN-IDI/2018/000174. The research of CS is supported by IISN-Belgium (convention 4.4503.15). CS is a Research Fellow of the F.R.S.-FNRS (Belgium).

\bibliography{references}

\end{document}